\documentstyle[prd,aps,epsfig,amssymb]{revtex}

\begin{document}
\newcommand{\bea}{\begin{eqnarray}}
\newcommand{\eea}{\end{eqnarray}}
\newcommand{\non}{\nonumber\\}
\newcommand{\al}{\alpha}
\newcommand{\be}{\beta}
\newcommand{\ld}{\lambda}
\draft
\preprint{\bf HUPD-0405\\
\bf hep-ph/}
\title{$B\to X_s\ell^+\ell^-$ and $B\to K\pi$ decays 
in vectorlike quark model}
\author{T. Morozumi $^1$\footnote{Email: morozumi@theo.phys.sci.hiroshima-u.ac.jp},  Z.H. Xiong $^1$\footnote{Email: xiongzh@theo.phys.sci.hiroshima-u.ac.jp}, T. Yoshikawa $^2$\footnote{Email: tasashi@eken.phys.nagoya-u.ac.jp}}
\address{$^1$ Graduate Scool of Science, Hiroshima University, 
Higashi-Hiroshima 739-8526, Japan}
\address{$^2$ Department of Physics, Nagoya University, Nagoya 464-8602, Japan}
\date{\today}

\maketitle

\begin{abstract}

 In the framework of SU(2) singlet down type vectorlike quark model,
we present a comprehensive analysis for decays
$B\to X_s\gamma,\ B\to X_sl^+l^-$ and $B\to K\pi$. 
As for $B\to X_s\gamma$, we  include
the QCD running from the mass of 
the down-type vector quark $D$ to weak scale in 
the scenario with the D quark much heavier than weak scale, and find 
that the running effect is small. Using the recent measurements 
of $ B\to X_sl^+l^-$, we extract rather stringent constraints 
on the size and CP violating phase  of $z_{sb}$, i.e., the tree level
FCNC coupling for $b\to s Z$. 
Within the bounds, we investigate various observables such as 
forward-backward asymmetry of $ b \to s l^+ l^-$, 
the decay rates of $B \to X_s \gamma$, and $B\to K\pi$.  
We find that (1) The forward-backward asymmetry may have 
large derivation from that of the SM  and is very sensitive to $z_{sb}$, 
and thus can be useful in probing the new physics. 
(2) By taking experimental errors at $2\sigma$ level, 
both experimental measurements 
for $B\to X_sl^+l^-$ and $B\to K\pi$ decays can be explained
in this model. 
\end{abstract}

\section{Introduction}
The study of flavor changing neutral currents (FCNC) in 
particle physics phenomenology has played a key role 
in advance of high energy physics in the past decades.
Due to the GIM mechanism, FCNC in the standard model (SM) 
arises only at higher loop level, and thus, it makes FCNC
phenomena a privileged ground to search for signs of
new physics beyond the SM.  However, in extension of 
the SM such as vector quark model (VQM), the CKM matrix 
is necessarily non-unitarity, leading to interaction 
$Z\bar{s}b$ at tree level, and hence potentially large new physics  
contributions can be expected. 

The rare radiative decays $B\to X_s\gamma$ and $B\to X_sl^+l^-$ 
are sensitive probes of new physics \cite{SB04}.
The branching ratio of the radiative decay $B\to X_s\gamma$ 
has been measured by  BaBar \cite{BaBar02}, CLEO\cite{CLEO01}, 
and ALEPH\cite{ALEPH98} and is in good agreement 
with the SM  predictions \cite{Chetyrkin97,Buras98}. 
Recently, the  rare decays $B\to X_sl^+l^-\ (l=e,\mu)$ 
also have been measured by BaBar\cite{Babar03} and Belle\cite{Belle03}. 
The average value is \cite{Nakao03} 
\bea
{\cal B}r^{ex}(B\to X_sl^+l^-)
=\left(6.2\pm 1.1^{+1.6}_{-1.3}\right)\times 10^{-6}.
\label{exbsll}
\eea
Although the process $B\to X_s\gamma$ constrains 
the parameters of the VQM \cite{Handoko95,Chang00}, since 
vectorlike down-type quark contributions to
$b\to s\gamma$ just occur at loop level as the case of the SM, 
the constraints on  $z_{sb}$, the tree level
FCNC coupling for $b\to s Z$, from $B\to X_s\gamma$ 
are less restrictive compared to those from 
processes governed by $b\to s l^+l^-$ transition. 

There are some studies regarding the constraints on model 
with extra singlet quark \cite{Handoko95,Chang00,Ahmady01,Barenboim01}.
In light of the improvement in the experimental data, 
it is necessary to present a comprehensive analysis in this model. 
Also, from the point of view of the model builder, it is important 
how the presence of the singlet quark may have impact on 
low energy phenomenology. In particular,
mass of the singlet quark, the coupling to ordinary quarks 
and weak gauge bosons are very important issues.  
We extend the previous studies and take the following points of
 the VQM into account:
 
(i) In the previous studies
\cite{Handoko95,Chang00,Ahmady01,Barenboim01,Aguilar02}, 
the  down-type vector quark $D$  is integrated out with $W,Z$ 
bosons and top quark together at $m_W$ scale,
neglecting the QCD running from $m_D$ to weak scale. 
In this work, we also 
consider the scenario with the D quark much heavier than weak scale.  
Firstly, the down-type vector quark is integrated out, generating an 
effective six-quark theory at $m_D$ scale. By using the renormalization 
group equation (RGE), the effective field theory is run down to the 
weak scale, at which the $W,Z$ bosons, Higgs and top quark are removed. 
Finally, the effective field theory is running down to the $b$ 
quark scale, as usual done in the SM.  From viewpoint of the theory, 
if there are different scales in a model,  including QCD running from 
heavier scale down to lighter one is important.

(ii) For inclusive decay $B\to X_sl^+l^-$, we 
include the four-quark operator contributions 
to one-loop matrix elements of operator ${\cal O}_9$
due to the tree level $Z\bar{s}b$ interaction.
We also consider the long distance contribution from resonance $\phi$. 
This is because the tree level FCNC $b \to s \bar{s} s$ generates the
new decay chain $B \to X_s \phi \to X_s l^+ l^-$. 
We show that
the dilepton mass distribution can be affected.

(iii) With the various improved treatments in  both short and long-distance
 contributions at hand,
 we obtain the rather stringent bound on the tree level Z FCNC coupling 
and its CP violating phase.  Within this bound, we study various observables
such as forward-backward asymmetry of $ b \to s l^+ l^-$, the decay rate of
$B \to X_s \gamma$.

(iv). The large 
electroweak penguin contribution to $b\to s$ transition has been
suggested in the present data of $B\to K\pi$
\cite{GNK,Yoshikawa03,Gronau-Rosner,BFRS,CL}.
In this work, we use factorization approach 
and study whether the large electroweak 
penguin contribution in $B\to K\pi$ can be explained or not
within FCNC constrained by rare decays $b\to sl^+l^-$.

This paper is organized as follows: In Section II we give a brief
description of the VQM. We present calculation
of $b\to s\gamma,\ b\to sl^+l^-$ transition in the VQM, 
including QCD running from
down-type vector quark mass scale to weak scale in 
Section III. 
Some new operators 
are introduced and the contributions of
electroweak penguin operators are taken into account. 
In Section IV , we extract constraints on 
size and phase of $z_{sb}$ from $B\to X_sl^+l^-$ experimental 
measurements. Using the bounds in Section  IV, 
we evaluate the forward-backward (FB) asymmetry 
and its zero-point $s_0$ for the rare B dileptonic decay,
as well as the branching ratio of $B\to X_s\gamma$, 
and show how they are affected by the new physics.
Section VI contributes to the study of $B\to K\pi$ decays. 
We found  both experimental measurements 
for $B\to X_sl^+l^-$ and $B\to K\pi$ decays can be explained
within the framework of vector-like quark model.
The anomalous dimension matrices needed in solving 
Wilson coefficients and all loop functions 
are collected in Appendix A and Appendix B, respectively. 

\section{Vector Quark Model}

In this section, we summarize the parameterization of quark mixings
in vectorlike quark model.  We focus on the model 
including  a singlet vectorlike down-type quark denoted by $D$, 
added to the standard model.

The difference between the new quark and ordinary quarks of the 
three SM generations is that, unlike the latter ones, both left-
and right-handed components of the former quark is  SU(2) 
singlet. Then the CKM matrix $V_{CKM}$ is enlarged to $3\times 4$ and 
can be expressed as \cite{Handoko95}
\bea
V_{CKM}^{j\be}\equiv \sum\limits_{i=1}^3 U^{ji}V^{\be i \ast},
\eea
where $U$, $V$ are $3\times 3$ and $4\times 4$ unitary matrices 
which relate the weak-eigenstates $\tilde{q}_L$ to
mass-eigenstates $q_L\ (q=u,d)$,
\bea
d_L^\be\equiv V^{\be\al}{\tilde d}_L^\al, \ \  \
u_L^i\equiv U^{ji}{\tilde u}_L^i.
\eea
The $4\times 4$ matrix $V$ covers ordinary ($\al=1,2,3$) and
and vectorlike quark ($\al=4$). 

The fact that the vectorlike quark  is isosinglet, leads to 
non-unitarity of the mixing matrix $V$ as
\bea 
z_{\al\be}&&\equiv\sum\limits_{i=1}^3V^{\al i}V^{\be i*}
=\sum\limits_{i=1}^3V_{CKM}^{i \al *}V_{CKM}^{ i \be} 
=\delta_{\al\be}-V^{\al 4}V^{\be 4*}.
\label{zsb}
\eea 
Geometrically, Eq.(\ref{zsb})  for $i=\alpha\neq \beta=j$ shows 
the quadrangle in the complex plane \cite{Branco93}. 
The deviations from the standard unitary triangles $(z_{ij}\neq 0)$ are 
going to vanish as the down type singlet mass (M) increases 
compared  with electroweak breaking scale $v$. 

The interaction Lagrangian for the quarks with the $W^{\pm}$ 
and Goldstone bosons $\chi^\pm$ reads
\bea
{\cal L}_{CC}&=&\frac{g}{\sqrt{2}m_W}V_{CKM}^{j\be}
\overline{u^j}\left(m_{u^j}L-m_{d^\be}R\right) d^\be \chi^+
+\frac{g}{\sqrt{2}}V_{CKM}^{j\be}
\overline{u^j}\gamma^\mu L d^\be W_\mu^+ +h.c.,
\eea
while the interaction Lagrangian for the down-type quarks with the $Z$, 
Higgs $H^0$ and Goldstone bosons $\chi^0$ is given by \cite{Handoko95}
\bea
{\cal L}_{NC}&=&-\frac{g}{\cos\theta_W}\sum\limits_{\al,\be}
\overline{d^\al}\gamma^\mu\left[\frac{1}{2}z^{\al\be}L
+e_d\sin^2\theta_W\delta_{\al\be}\right]d^\be Z_\mu
-\frac{g}{2m_W}\sum\limits_{\al,\be}
\overline{d^\al}z_{\al\be}\left[m_{d^\al}L+m_{d^\be}R\right]d^\be H^0\non
&&+i\frac{g}{2m_W}\sum\limits_{\al,\be}
\overline{d^\al}z_{\al\be}\left[m_{d^\al}L-m_{d^\be}R\right]d^\be \chi^0,
\eea
where $e_d=-1/3$ is the electric charge of the down-type quark.

So far, the results are general. To discuss the effects of singlet
quark on unitarity of the CKM matrix, we must keep the length 
of the sides of quadrangles to the order of $\frac{v^2}{M^2}$. Then 
we can discuss the detailed structure of the quadrangles.
For that purpose, we devise the parameterization of $3\times 4$
non-unitarity matrix based upon the systematic expansion of 
$\frac{v}{M}$.  Using the   parameterization, we show 
the quadrangle of $b\bar{s}$ sector, which will be used in later section.

It is completely general to start with mass matrix of the down type 
quarks as  follows\cite{Branco93}:
\bea
{\cal M}_d\equiv\left(\begin{array}{cc}
m_0 &J\\
0 & M\end{array}\right), 
\eea
with charged current
\bea
\bar{u}_L\gamma_\mu K^0d_L,\nonumber
\eea
where $m_0$ is a real diagonal $3\times 3$ matrix with $(m_0)_{ij}=
m_{0i}\delta_{ij}$. $J$ is a  $3\times 1$
matrix, $J^T=(J_1,J_2,J_3)$, and $J_1$ is real 
while $J_2$ and $J_3$ are complex.
$M$ is a singlet quark mass parameter and can be taken as real.      
$K^0$ is a $3\times 3$ unitary matrix and can be  parameterized 
by $\lambda,\ A,\ \rho,\ \eta$ as in the CKM matrix of the SM. Note that
due to non-vanishing Z FCNC couplings, the  values of 
$\lambda,\ A,\ \rho,\ \eta$ can be different from those of the SM.  
The matrix ${\cal M}_d{\cal M}_d^\dag$ can be diagonalized by a $4\times 4$
unitary matrix $W$, 
\bea
W=\left(\begin{array}{cc}
\Omega &R\\
S &T
\end{array}\right),\ \  
W{\cal M}_d{\cal M}_d^\dag W^\dag=\left(\begin{array}{cc}
\bar{m}^2 &0\\
0 &\bar{M}^2\end{array}\right).
\eea
Thus,   the $3\times 4$
non-unitary matrix $V_{CKM}$ is a submatrix of $W$, $V_{CKM}=(\Omega, 
R)$. 
The matrices $\Omega $ and $S$ satisfy following equations:
\bea
(m_0^2+JJ^\dag)\Omega +MJS=\Omega\bar{m}^2,\non
MJ^\dag \Omega+M^2S=S\bar{M}^2,\non
\Omega\Omega^\dag+RR^\dag=1. 
\eea
By eliminating $S$ and using $R\simeq \frac{J}{M}$, 
we obtain $\Omega$ to the order of 
$\frac{JJ^\dag}{M^2}$ as
\bea
\Omega =\left(\begin{array}{ccc}
1-\frac{\Delta_{11}}{2}&\Delta_{12}\frac{m_{02}^2}{\Delta m^2_{012}}
&\Delta_{13}\frac{m_{03}^2}{\Delta m^2_{013}}\\
\Delta_{21}\frac{m_{01}^2}{\Delta m^2_{021}}&1-\frac{\Delta_{22}}{2}&
\Delta_{23}\frac{m_{03}^2}{\Delta m^2_{023}}\\
\Delta_{31}\frac{m_{01}^2}{\Delta m^2_{031}}&
\Delta_{32}\frac{m_{02}^2}{\Delta m^2_{032}}&1-\frac{\Delta_{33}}{2}
\end{array}\right),
\label{KK}
\eea
where
\bea
\Delta_{ij}=\frac{J_iJ^*_j}{M^2}, \ \ 
\Delta m^2_{0ij}=m^2_{0i}-m^2_{0j}.
\eea
Now we can contact the expression of Z coupling $z_{ij}$ with 
$\Delta_{ij}$ in this approximation. From Eq.(\ref{zsb}) and 
(\ref{KK}), we obtain 
\bea
z_{ij}=\delta_{ij}-\Delta_{ij}.
\eea
Finally, considering $m_{01}\simeq m_d\ll m_{02}\simeq m_s\ll m_{03}\simeq m_b$, we have 
\bea
V_{CKM}\simeq K^0\left(\begin{array}{cccc}
1-\frac{\Delta_{11}}{2}
&z_{ds}
&z_{db}& \sqrt{\Delta_{11}}\\
0&1-\frac{\Delta_{22}}{2}
&z_{sb}&\sqrt{\Delta_{22}}e^{i\delta_2}\\
0&0&1-\frac{\Delta_{33}}{2}
&\sqrt{\Delta_{33}}e^{i\delta_3}
\end{array}\right),
\label{parCKM}
\eea
where $\delta_i=arg(J_i)\ (i=2,3)$. We note that there are three independent CP
violating phases.
Thus, by parameterizing the CKM matrix, we link
the CKM matrix $V_{CKM}$ to  
a unitarity matrix $K^0$ and $Z$ FCNC couplings. 

\begin{figure}[htb]
\begin{center}
\epsfig{file=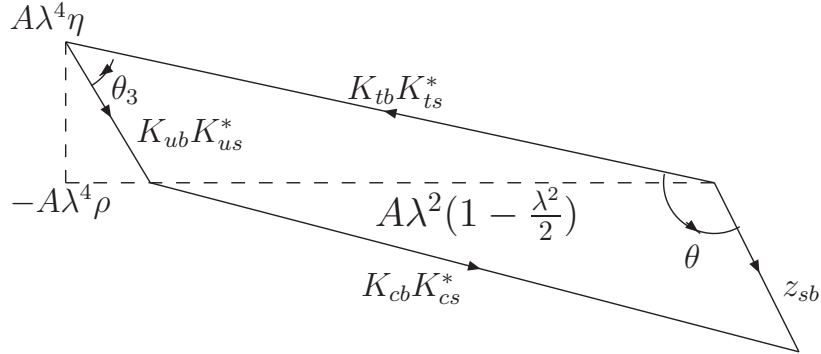,width=12cm}
\end{center}
\caption{ The quadrangle in  $b\bar{s}$ sector.}
\label{XX}
\end{figure}

Using the experimental measurements for $B_d\bar{B}_d$, $K\bar{K}$ 
mixings, CP asymmetry of $B\to \Psi K_S$ and CKM matrix elements, 
we can constrain the  Z FCNC in  $b\bar{d}$, $s\bar{d}$  sectors, 
and investigate the correlations among the Z FCNC in $b\bar{d}$, $s\bar{d}, 
b\bar{s}$ sectors.
The detailed study of them will be presented elsewhere. 
In this work, we will focus attention on the Z FCNC in $b\bar{s}$ sector,
 assuming that the Z FCNC effects on mixings and decays  of $K,\ B_d$
are negligible. In this approximations, we obtained
\bea
K_{ub}K_{us}^* &=&A\lambda^4(\rho-i\eta),\non  
K_{cb}K_{cs}^* &=&A\lambda^2(1-\frac{\lambda^2}{2})+z_{sb},\non 
K_{tb}K_{ts}^* &=&-A\lambda^2(1-\frac{\lambda^2}{2})
-A\lambda^4(\rho-i\eta),  
\eea
where  $K_{ij}\equiv (V_{CKM})_{ij}$ for $i,j=1,2,3$. 
The corresponding quadrangle in  $b\bar{s}$ sector
is shown in Fig. \ref{XX}. 

\section{$b\to s\gamma$ and $b\to s \ell^+\ell^-$ transitions in VQM}

In VQM, the down-type vector quark 
may be much heavier than weak scale.  In a theory with
different mass scales, the heavier scale should be integrated out 
firstly, then Wilson coefficients are run
from heavier scale to  low scale by using renormalization group 
equation.  Only in case of  $m_D$ is about the weak scale, 
can $W,Z$ boson, Higgs boson and top quark be integrated out 
together. In this work, we consider two possibilities as follows:

\subsection {{\em Scenario A}:  $\delta_D=\frac{m^2_{Z,W,t,H}}{m_D^2}\ll 1$}
 
In this scenario,  we first integrate out the heavy $D$ quark, 
introducing dimension-5 and dimension-6 operators.
By keeping only leading order terms of $\delta_D$, 
we obtain the effective Hamiltonian for $b\to s\gamma^*(g^*)$ as: 
\bea
{\cal H}^{new}_{eff}(b\to \gamma^*(g^*))
=\frac{4G_F}{\sqrt{2}}z_{4s}^*z_{4b}\sum_iC_i(\mu)O_i.
\label{opernew}
\eea
A complete basis for the local  operators is listed below:
\bea
O_{LR}^1&=&-\frac{1}{16\pi^2}m_b\bar{s}_L{\cal D}^2b_R,\non
O_{LR}^2&=&\frac{1}{16\pi^2}\mu^{\epsilon/2} g_s
m_b\bar{s}_L\sigma^{\mu\nu}T^ab_RG_{\mu\nu},\non
O_{LR}^3&=&\frac{1}{16\pi^2}\mu^{\epsilon/2} ee_d
m_b\bar{s}_L\sigma^{\mu\nu}b_RF_{\mu\nu},\non
Q_{LR}^H&=&\frac{1}{2}\mu^{\epsilon}g_s^2 m_bH^0H^0\bar{s}_Lb_R,\non
Q_{LR}^\chi&=&\frac{1}{2}\mu^{\epsilon}g_s^2 m_b\chi^0\chi^0\bar{s}_Lb_R,\non
P_{L}^{1,A}&=&-i\frac{1}{16\pi^2}\bar{s}_LT_{\mu\nu\sigma}^A
{\cal D}^\mu {\cal D}^\nu{\cal D}^\sigma b_R,\non
P_{L}^{2}&=&\frac{1}{16\pi^2}\mu^{\epsilon}
\bar{s}_L\gamma^{\mu}b_R\partial^\nu F_{\mu\nu},\non
R_L^{1,H}&=&i\frac{1}{2}\mu^{\epsilon}g_s^2 H^0H^0\bar{s}_L\not {\cal
D}b_L,\non
R_L^{1,\chi}&=&i\frac{1}{2}\mu^{\epsilon}g_s^2 \chi^0\chi^0\bar{s}_L\not {\cal
D}b_L,\non
R_L^{2,H}&=&i\mu^{\epsilon}g_s^2 ({\partial}^\sigma H^0)H^0
\bar{s}_L\gamma_\sigma b_L,\non
R_L^{2,\chi}&=&i\mu^{\epsilon}g_s^2 ({\partial}^\sigma \chi^0)\chi^0
\bar{s}_L\gamma_\sigma b_L,
\label{operatord}
\eea
where $T^a$ stand for $SU(3)_{color}$ generators, 
$F_{\mu\nu}$ and $G_{\mu\nu}$ are field strength of
photon and gluon respectively. 
The covariant derivative is given by 
\bea
{\cal D}_\mu=\partial_\mu-i\mu^{\epsilon/2}g_sG_\mu^aT^a
-i\mu^{\epsilon/2}ee_dA_\mu,
\eea
with $\epsilon=4-d$.
The tensor $T_{\mu\nu\sigma}^A\ (A=1,2,3, 4)$ appearing 
in $P_L^{1,A}$ have the following Lorentz structures
\cite{Cho91,Gao95}:
\bea
T_{\mu\nu\sigma}^1=g_{\mu\nu}\gamma_{\sigma}, \ \ \ 
T_{\mu\nu\sigma}^2=g_{\mu\sigma}\gamma_{\nu},\ \ \ 
T_{\mu\nu\sigma}^3=g_{\nu\sigma}\gamma_{\mu}, \ \ \
T_{\mu\nu\sigma}^4=-i\epsilon_{\mu\nu\sigma\tau}\gamma^\tau\gamma_5.
\eea

\begin{figure}[htb]
\centerline{\epsfig{file=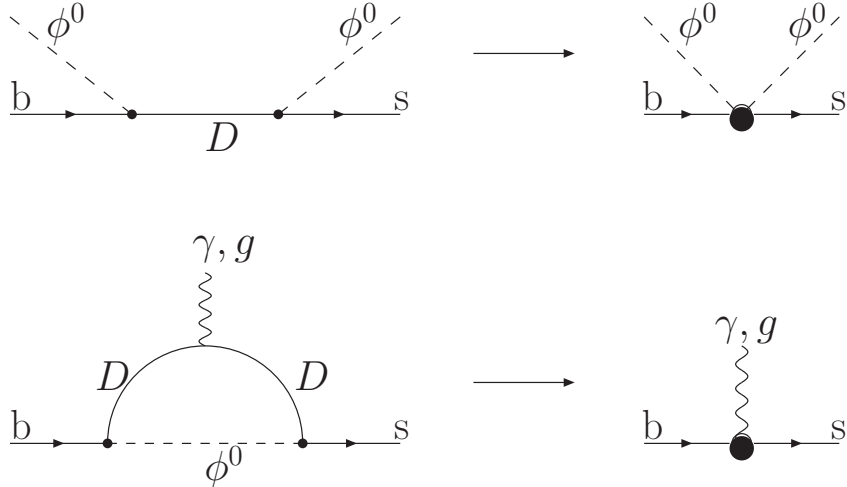,width=12cm}}
\caption{Matching conditions at scale $m_D$ in full theory
(left) and in the intermediated effective field theory (right).
$\phi^0$ can be  $H^0$ and $\chi^0$.}
\label{matchD}
\end{figure}
To match the full theory into an effective theory,
the diagrams to be matched  at $m_D$ scale are displayed in Fig. \ref{matchD}.
To determine the coefficients of the $\Gamma^{bs\phi\phi}$,
$\Gamma^{bs\gamma\phi\phi}$,
$\Gamma^{bs g\phi\phi}\ (\phi=H^0, \chi^0)$ at scale $m_D$, we just 
need to match the full theory with effective theory at tree level.
We obtain 
\bea
C_{Q_{LR}^H}(m_D)=-C_{Q_{LR}^\chi}(m_D)=-\frac{1}{g_s^2},\\
C_{R_L^{1,H}}(m_D)=2C_{R_L^{2,H}}(m_D)=-\frac{1}{2g_s^2},\\
C_{R_L^{1,\chi}}(m_D)=2C_{R_L^{2,\chi}}(m_D)=-\frac{1}{2g_s^2}.
\eea

Other coefficients can be determined by matching one loop diagrams shown 
in Fig. \ref{matchD}. After straightforward calculations, we have  
\bea
C_{O_{LR}^i}(m_D)&=&0 \ \ (i=1,2,3),\non
C_{P_L^{1,1}}(m_D)&=&C_{P_L^{1,3}}(m_D)=-\frac{11}{18},\non
C_{P_L^{1,2}}(m_D)&=&\frac{8}{9},\non
C_{P_L^{1,4}}(m_D)&=&-\frac{1}{2},\non
C_{P_L}^2(m_D)&=&0.
\eea
The values  are understood as sum of $H^0,\ \chi^0$ contributions. 
Cancellation between Higgs and would-like Goldstone boson 
$\chi^0$  leads to  $C_{O_{LR}}^i(m_D)=0\ (i=1,2,3)$ in leading order. 

The running of Wilson coefficients $C_i(\mu)$ from $m_D$ down to 
weak scale is governed by anomalous dimension $\gamma$ through RGE
\bea
\mu\frac{d}{d\mu}C_i(\mu)=\sum_j(\gamma^T)_{ij}C_j(\mu).
\label{REQD}
\eea

We calculate one-loop diagrams with operator
insertions and present the anomalous dimensions $\gamma$ in Appendix A. 
Using the anomalous dimensions (\ref{DM1}) and (\ref{DM2}), we can solve
the RGE (\ref{REQD}) and have the coefficients of the operators at
weak scale $m_W^+$. They are given by,
\bea
C_{O_{LR}^1}(m_W^+)&=&\frac{247}{548}\zeta^{-\frac{4}{21}}
                 +\frac{336}{8905}\zeta^{\frac{113}{126}}
-\frac{551}{780}\zeta^{\frac{8}{21}}
                 +\frac{1}{6}\zeta^{\frac{2}{3}},\non
C_{O_{LR}^2}(m_W^+)&=&\frac{247}{1096}\zeta^{-\frac{4}{21}}
                 +\frac{168}{8905}\zeta^{\frac{113}{126}}
                 -\frac{223}{780}\zeta^{\frac{8}{21}}
                 +\frac{1}{24}\zeta^{\frac{2}{3}},\non
C_{O_{LR}^3}(m_W^+)&=&\frac{247}{1096}\zeta^{-\frac{4}{21}}
                 +\frac{168}{8905}\zeta^{\frac{113}{126}}
                 -\frac{223}{780}\zeta^{\frac{8}{21}}
                 +\frac{5}{12}\zeta^{\frac{2}{3}}
                 -\frac{3}{8}\zeta^{\frac{16}{21}},\non
C_{P_L^{1,1}}(m_W^+)&=&C_{P_L^{1,3}}(m_W^+)=-\frac{247}{548}
\zeta^{-\frac{4}{21}}-\frac{791}{4932}\zeta^{\frac{113}{126}},\non
C_{P_L^{1,2}}(m_W^+)&=&\frac{247}{548}\zeta^{-\frac{4}{21}}
                    +\frac{1}{12}\zeta^{\frac{8}{21}}
+\frac{875}{2466}\zeta^{\frac{113}{63}},\non
C_{P_L^{1,4}}(m_W^+)&=&-\frac{247}{598}\zeta^{-\frac{8}{21}}
-\frac{1}{12}\zeta^{\frac{8}{21}}
+\frac{14}{411}\zeta^{\frac{113}{126}},\non
C_{P_L^2}(m_W^+)&=&0,
\eea
where $\zeta=\alpha_s(m_D)/\alpha_s(m_W^+)$.

In order to continue running the basis operator coefficients 
from  $m_W$ scale down to $b$ quark scale, we use the  
effective QCD-corrected Hamiltonian obtained by integrating out 
the $W,Z$ bosons, would-be Goldstone boson, Higgs boson and 
top quark. The effective Hamiltonian describing $b\to sl^+l^-$ 
transition  reads \cite{Chetyrkin97,Ciuchini94}
\bea
{\cal H}_{eff}(b\to sl^+l^-)
&=&-\frac{4G_F}{\sqrt{2}}K_{tb} K^*_{ts}
\left[\sum\limits_{i=1}^{10}{\widetilde C}_i(\mu){\cal Q}_i
+\widetilde{C}_{7\gamma}(\mu){\cal Q}_{7\gamma}
+\widetilde{C}_{8G}(\mu){\cal Q}_{8G}
+{\cal C}_{9}(\mu){\cal O}_{9}
+{\cal C}_{10}(\mu){\cal O}_{10}\right],
\label{heff}
\eea
where
\bea
{\cal Q}_1&=&(\bar{s}_i\gamma^\mu L c_j)(\bar{c}_j\gamma_\mu Lb_i),\non
{\cal Q}_2&=&(\bar{s}\gamma^\mu L c)(\bar{c}\gamma_\mu L b),\non
{\cal Q}_3&=&(\bar{s}\gamma^\mu L b)\sum_{q}(\bar{q}\gamma_\mu L q),\non
{\cal Q}_4&=&(\bar{s}_i\gamma^\mu L  b_j)\sum_{q}(\bar{q}_j\gamma_\mu Lq_i),\non
{\cal Q}_5&=&(\bar{s}\gamma^\mu L b)\sum_{q}(\bar{q}\gamma_\mu R q),\non
{\cal Q}_6&=&(\bar{s}_i\gamma^\mu L b_j)\sum_{q}(\bar{q}_j\gamma_\mu R q_i),\non
{\cal Q}_7&=&\frac{3}{2}(\bar{s}\gamma^\mu L b)\sum_q e_q (\bar{q}
\gamma_\mu R q),\non
{\cal Q}_8&=&\frac{3}{2}(\bar{s}_i\gamma^\mu L b_j)\sum_q e_q 
(\bar{q}_j\gamma_\mu R q_i),\non
{\cal Q}_9&=&\frac{3}{2}(\bar{s}\gamma^\mu L b)
\sum_q e_q (\bar{q}\gamma_\mu Lq),\non
{\cal Q}_{10}&=&\frac{3}{2}(\bar{s}_i\gamma^\mu L b_j)
\sum_q e_q (\bar{q}_j\gamma_\mu Lq_i),\non
{\cal Q}_{7\gamma}&=&\frac{e}{16\pi^2}m_b\bar{s}_i\sigma^{\mu\nu}R
b_i F_{\mu\nu},\non
{\cal
Q}_{8G}&=&\frac{g_s}{16\pi^2}\bar{s}_i\sigma^{\mu\nu}RT_{ij}^ab_jG^a_{\mu\nu},
\non
{\cal O}_{9}&=&\frac{e^2}{16\pi^2}(\bar{s}\gamma^\mu L b)
(\bar{l}\gamma_\mu l),\non
{\cal O}_{10}&=&\frac{e^2}{16\pi^2}(\bar{s}\gamma^\mu L b)
(\bar{l}\gamma_\mu\gamma_5 l).
\label{operators}
\eea

To match the operator set in (\ref{operatord}) onto these operators,
we use the equations of motion to reduce all remaining two-quark 
operators to the gluon and photon magnetic moment operators $O_{LR}^2$
and $O_{LR}^3$ which are redefined as ${\cal Q}_{8G}$ and 
${\cal Q}_{7\gamma}$ in (\ref{operators}).

\begin{figure}[htb]
\centerline{\epsfig{file=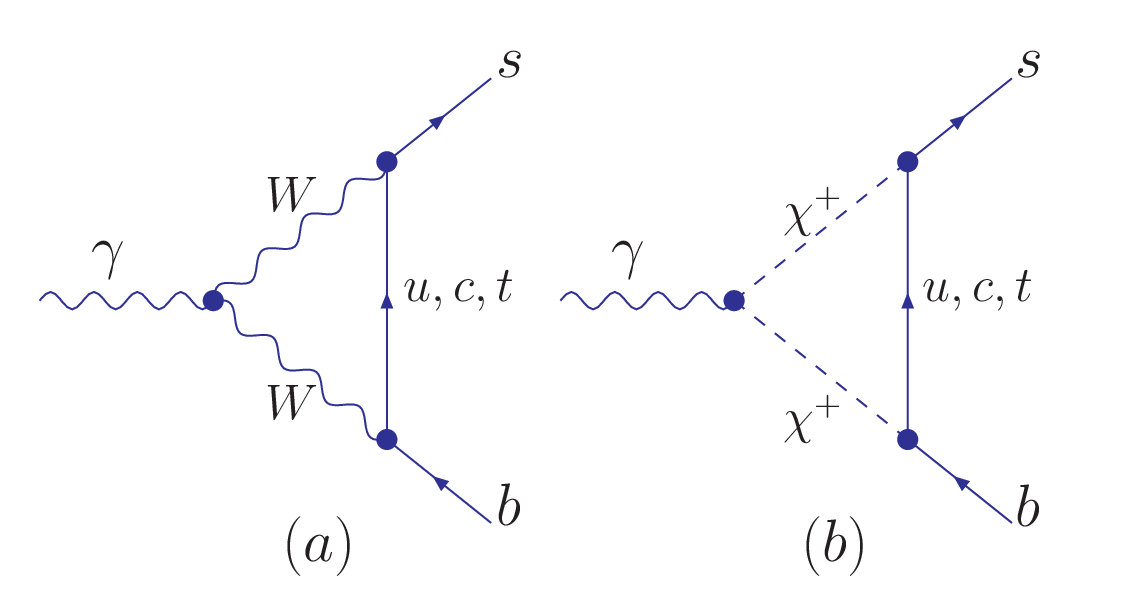,width=7.5cm}
\epsfig{file=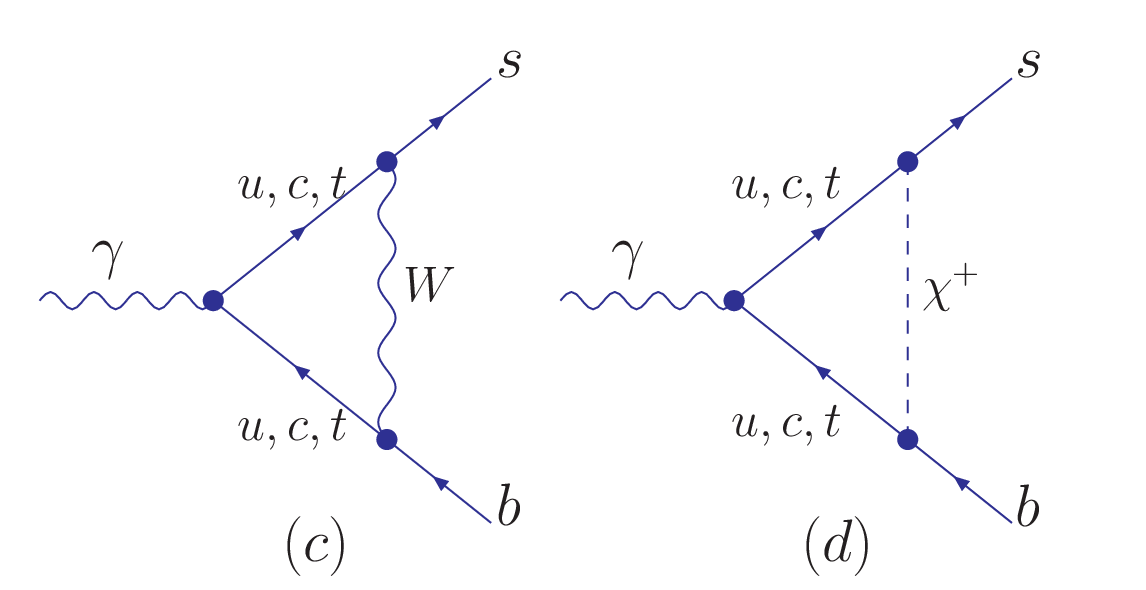,width=7.5cm}}
\caption{Charged boson mediated penguin 
diagram contributing to $b\to s\gamma$.}
\label{Fig:Fig1}
\end{figure}

Firstly, we present the values of the Wilson coefficients
at $m_W$ scale  calculated in NDR and HV 
schemes  as follows with small $V_{ub}V_{us}^*$ omitted:
\bea
{\widetilde C}_1(m_W)&=&\frac{11}{2}(1+2\xi)\frac{\alpha_s(m_W)}{4\pi},\non
{\widetilde C}_2(m_W)&=&1-\frac{z_{sb}}{K_{tb}K_{ts}^*}
-\frac{11}{6}(1+2\xi)\frac{\alpha_s(m_W)}{4\pi}
-\frac{35}{18}\frac{\alpha_{em}}{4\pi},\non
{\widetilde C}_3(m_W)&=&-\frac{1}{6}
\frac{z_{sb}}{K_{tb}K_{ts}^*}-
\frac{1}{6}\frac{\alpha_s(m_W)}{4\pi}
\left[E(x_t)+\frac{2}{3}\xi\right]
+\frac{\alpha_{em}}{6\pi}\frac{2B(x_t)+C(x_t)}{\sin^2\theta_W}
,\non
{\widetilde C}_4(m_W)&=&
-3{\widetilde C}_5(m_W)={\widetilde C}_6(m_W)=-\frac{1}{2}
\left[E(x_t)+\frac{2}{3}\xi\right]\frac{\alpha_s(m_W)}{4\pi},\non
{\widetilde C}_7(m_W)&=&-\frac{2}{3}\sin^2\theta_W\frac{z_{sb}}{K_{tb}K_{ts}^*}
+\frac{\alpha_{em}}{6\pi}\left[4C(x_t)+D(x_t)-\frac{4}{9}(1+\xi)\right]
,\non
{\widetilde C}_8(m_W)&=&{\widetilde C}_{10}(m_W)=0,\non
{\widetilde C}_9(m_W)&=&-\frac{2}{3}\cos^2\theta_W\frac{z_{sb}}{K_{tb}K_{ts}^*}
+\frac{\alpha_{em}}{6\pi}
\left[\frac{10B(x_t)-4C(x_t)}{\sin^2\theta_W}
+4C(x_t)+D(x_t)-\frac{4}{9}(1+\xi)\right],\non
{\widetilde C}_{7\gamma}(m_W)&=&-\frac{1}{2}A(x_t)
+\frac{z_{sb}}{K_{tb}K_{ts^*}}\left[\frac{23}{36}
-\frac{1}{2}C_{O_{LR}^1}(m_W^+)+C_{O_{LR}^3}(m_W^+)
-\frac{1}{4}C_{P_L^{1,2}}(m_W^+)
+\frac{1}{4}C_{P_L^{1,4}}\right.\non
&&\left.+e_d\left(\frac{1}{3}+\frac{1}{9}\sin^2\theta_W\right)\right]\\
{\widetilde C}_{8G}(m_W)&=&-\frac{1}{2}F(x_t)
+\frac{z_{sb}}{K_{tb}K_{ts^*}}\left[
\frac{1}{3}-\frac{1}{2}C_{O_{LR}^1}
(m_W^+)+C_{O_{LR}^2}(m_W^+)
-\frac{1}{2}C_{P_L^{1,1}}(m_W^+)
-\frac{1}{4}C_{P_L^{1,2}}(m_W^+)
\right.\non
&&\left.+\frac{1}{4}C_{P_L^{1,4}}
-3e_d\left(\frac{1}{3}+\frac{1}{9}\sin^2\theta_W\right)\right],\non
{\cal C}_9(m_W)&=&\frac{\pi}{\alpha_{em}}\frac{z_{sb}}{K_{tb}K_{ts}^*}
(-1+4\sin^2\theta_W)+\frac{C(x_t)-B(x_t)}{\sin^2\theta_W}-4C(x_t)-D(x_t)
+\frac{4}{9}(1+\xi)+O\left(\frac{z_{sb}}{K_{tb}K_{ts}^*}\right),\non
{\cal C}_{10}(m_W)&=&\frac{\pi}{\alpha_{em}}
\frac{z_{sb}}{K_{tb}K_{ts}^*}+\frac{B(x_t)-C(x_t)}{\sin^2\theta_W}
+O\left(\frac{z_{sb}}{K_{tb}K_{ts}^*}\right),
\label{cw}
\eea
where $\xi=0$ in NDR scheme and $\xi=-1$ in HV
scheme. The loop functions can be found in Appendix B. 

\begin{figure}[htb]
\centerline{\epsfig{file=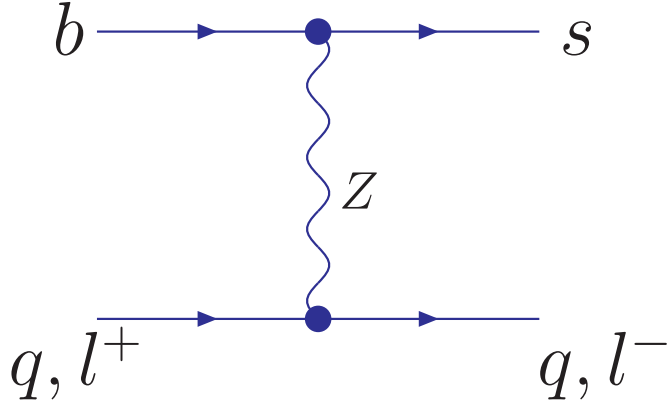,width=10cm}}
\caption{Tree level Feynman diagram 
contributing to $b\to s\gamma$ and $b\to sl^+l^-$.}
\label{Fig:Fig3}
\end{figure}

\begin{figure}[htb]
\begin{center}
\centerline {\epsfig{file=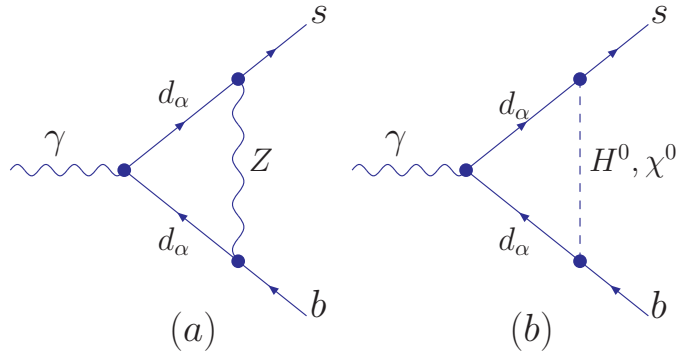,width=10cm}}
\caption{Neutral boson mediated penguin 
diagram contributing to $b\to s\gamma$.}
\label{Fig:Fig2}
\end{center}
\end{figure}

The terms proportional to $\frac{z_{sb}}{K_{tb}K_{ts}^*}$  in 
${\widetilde C}_{3,7,9}(m_W)$ and the first term of 
${\cal C}_{9,10}(m_W)$ come from the tree-level diagram
as displayed in Fig. \ref{Fig:Fig3}, 
the term proportional to $\frac{z_{sb}}{K_{tb}K_{ts}^*}$  
in  ${\widetilde C}_{2}(m_W)$ comes from tree diagram due to 
the non-unitarity of CKM matrix in VQM.  
In expressions of ${\widetilde C}_{7\gamma,8G}(m_W)$,  
the first constant terms proportional to 
$\frac{z_{sb}}{K_{tb}K_{ts}^*}$ 
come from the charged current  one-loop diagram Fig.\ \ref{Fig:Fig1}
due to the non-unitarity of CKM matrix in VQM, 
other  terms proportional to $\frac{z_{sb}}{K_{tb}K_{ts}^*}$
come from the neutral current  one-loop diagrams
shown in Fig. \ref{Fig:Fig2}.
As in expressions  ${\cal C}_{9,10}(m_W)$, the 
second  terms proportional to $\frac{z_{sb}}{K_{tb}K_{ts}^*}$
denoted as $O\left(\frac{z_{sb}}{K_{tb}K_{ts}^*}\right)$
come from the charged and  neutral current one-loop diagrams.
In deriving above equation, we have used the unitarity relation 
\bea
z_{4b}z_{4s}^*=-z_{sb}
\eea  
which is a direct result of Eq. (\ref{zsb}). 

At this moment, we would like to point out that  
the contributions from the $Z$-penguin charged current one-loop diagrams 
in the VQM to ${\cal C}_{9,10}(m_W)$, have divergent terms due to the 
non-unitarity of CKM matrix. Although these divergences 
can be removed by renormalizing the tree level FCNC which 
exists in VQM Lagrangian, they are scheme dependent.   
However, these scheme dependences are  subleading  
compared to the first terms of ${\cal C}_{9,10}(m_W)$, which
we will neglect them in our calculation. Under this approximation they are
not relevant for the inclusive dileptonic decays.

\subsection{{\em Scenario B}: $\delta_D=\frac{m^2_{Z,W,t,H}}{m_D^2}\sim  1$}

In this scenario, the top and $D$ quark, W and  Z bosons
can be integrated out together. The corresponding initial 
values of Wilson  coefficients ${\widetilde C}_{7\gamma,8G}(m_W)$
are changed to \cite{Handoko95}
\bea
{\widetilde C}_{7\gamma}(m_W)&=&-\frac{1}{2}A(x_t)
+\frac{z_{sb}}{K_{tb}K_{ts}^*}\left[\frac{23}{36}+
\left(f_D^Z(y_D)+f_D^H(w_D)+f_D^\chi(y_D)\right)+
e_d\left(\frac{1}{3}+\frac{1}{9}\sin^2\theta_W\right)\right],\non
{\widetilde C}_{8G}(m_W)&=&-\frac{1}{2}F(x_t)
+\frac{z_{sb}}{K_{tb}K_{ts}^*}\left[\frac{1}{3}
-3\left(f_D^Z(y_D)+f_D^H(w_D)+f_D^\chi(y_D)\right)
-3e_d\left(\frac{1}{3}+\frac{1}{9}\sin^2\theta_W\right)\right].
\eea
where $y_D=m_D^2/m_Z^2,\  w_D=m_D^2/m_H^2$. 
Other Wilson coefficients are the same as  {\it  Scenario A } we  discussed. 
The functions $f_y^x$ stands for the  contribution from boson $x$ 
mediated penguin one-loop diagram with quark $y$ in loops and 
are presented in Appendix B.  
\begin{figure}[htb]
\centerline{\epsfig{file=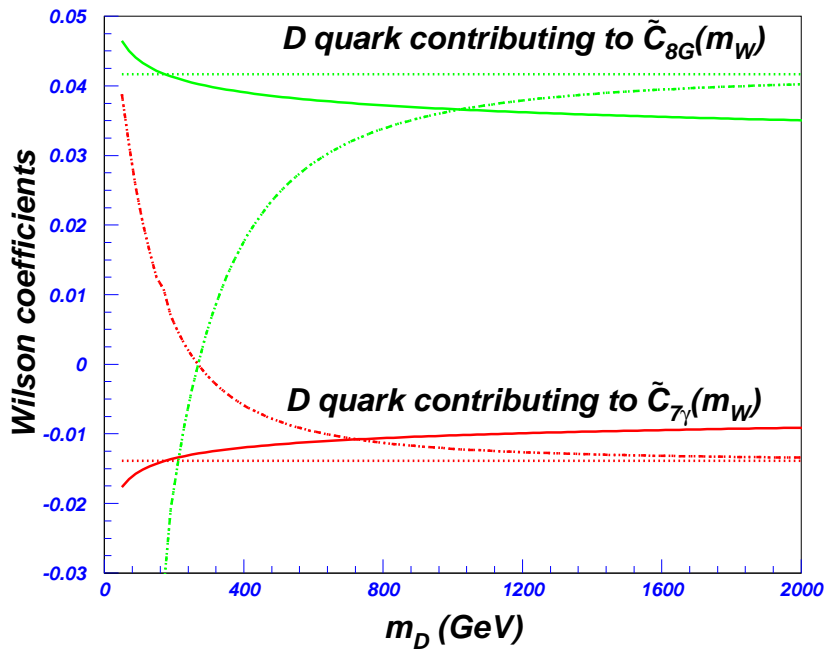,width=12cm}}
\caption{The down-type vector quark contributions to 
Wilson coefficients of ${\cal Q}_{7\gamma}$ 
and ${\cal Q}_{8G}$ at $m_W$ scale in unit of 
$\frac{z_{sb}}{K_{tb}K_{ts}^*}$\ $(m_H=m_t)$.  
The solid and dashed lines stand for the results in
{\em Scenario A} with or without consideration of 
 QCD running effect from $m_D$ to $m_W$, respectively. 
The corresponding values in {\em Scenario B} are denoted 
by dot-dashed lines.}
\label{figcnw}
\end{figure}

In  Fig.\ref{figcnw} we plot the down-type vector quark contribution to 
the Wilson coefficients ${\widetilde C}_{7\gamma}$ 
and ${\widetilde C}_{8G}$ as a function $m_D$ 
at $m_W$ scale, and demonstrate the QCD running effects 
from $m_D$  down to $m_W$ in {\em Scenario A}.

As a consistency check, we can  see that if the QCD correction 
is neglected by setting $\zeta=1$ and only leading order terms of 
$\delta_D$ are kept, for on-shell quarks and photon,   
our results ${\widetilde C}_{7\gamma}(m_W)$ and ${\widetilde C}_{8G}(m_W)$ 
in {\em Scenario A} would produce the results exactly in {\em Scenario B}.

At the end of this subsection, we
make some emphasis  on the results obtained by using different approaches 
as follows:

The effect of running from $m_D$ down to weak scale
on  the contributions from  neutral current mediated by 
$D$ quark in {\em Scenario A} is large, as shown in Fig.\ \ref{figcnw}.
However, at $m_W$ scale, since dominant new  contribution
comes from the charged current diagrams, the total results
of ${\widetilde C}_{7\gamma,8G}(m_W)$ are changed slightly. 
This indicates that  the dependence of the branching ratio of 
$B\to X_s\gamma$ on $m_D$ is quite weak in both {\em Scenarios}.
Therefore, we can not  extract the mass of 
down-type vector quark from this analysis.

\section{ Constraints on $z_{sb}$ in VQM from  inclusive decays 
$B\to X_s\ell^+\ell^-$}

\subsection{Solutions of Wilson coefficients}
\label{sec:bsll}

To obtain some predictions for inclusive $B$ rare decays, 
we need to determine  the Wilson coefficients at $m_b$ scale.
Expanding anomalous dimension matrix and Wilson coefficient as
\bea
\gamma&=&\gamma^{(0)}\frac{\alpha_s(\mu)}{4\pi}+\gamma^{(1)}(\frac{\alpha_s(\mu)}{4\pi})^2
+\cdots,\\
C(\mu)&=&C^{(0)}(\mu)+C^{(1)}(\mu)\frac{\alpha_s(m_W)}{4\pi}+\cdots,
\eea
we solve the Wilson coefficients up to order $\alpha_s(m_W)$
according to the RGE (\ref{REQD}).
The relevant $10\times 10$ one-loop anomalous dimension
matrix  among ${\cal Q}_{1-10}$,  
$\gamma^{(0)}$, and two-loop anomalous dimension matrices
among  ${\cal Q}_{1-10}$ with 
${\cal Q}_{7\gamma,8G}$ and ${\cal O}_{9}$ needed 
in our calculations are collected in Appendix A. 

Using the $10\times 10$ matrix $\gamma^{(0)}$ 
in Appendix, and initial values presented in Section II,
we obtain the solutions 
\bea
{\widetilde C}_i^{(0)}(\eta)&=&\sum\limits_{i,l=1}^{10}
{\hat V}^{-1}_{il}\eta^{a_l}{\hat V}_{l2}
-\frac{z_{sb}}{K_{tb}K_{ts}^*}{\hat V}^{-1}_{il}
\sum\limits_{l}^{10}
\eta^{a_l}\left[{\hat V}_{l2}
+\frac{1}{6}{\hat V}_{l3}+\frac{2}{3}\sin^2\theta_W{\hat V}_{l7}+
\frac{2}{3}\sin^2\theta_W{\hat V}_{l9}\right].
\label{cb0i}
\eea
where $\eta=\alpha_s(m_W)/\alpha_s(\mu),\ 
{\widetilde C}_k^{(0)}(1)\equiv {\widetilde C}_k^{(0)}(m_W)$. The eigenvalues $a_l$ are obtained by diagonalizing the $10\times 10$ anomalous dimension matrix
$\frac{\gamma^{(0)}}{2\beta_0}$,
\bea
({\hat V}\frac{\gamma^{(0)}}{2\beta_0}{\hat V}^{-1})_{lj}=a_l\delta_{lj}
\nonumber.
\eea

At leading order, the solution of ${\widetilde  C}_{7\gamma}(\eta)$ 
is given by
\bea
{\widetilde  C}_{7\gamma}(\eta)&=&
\eta^{\frac{\gamma_{\gamma\gamma}}{2\beta_0}}
{\widetilde  C}_{7\gamma}(m_W)
+\frac{\gamma_{G\gamma}}{\gamma_{GG}-\gamma_{\gamma\gamma}}
\left(\eta^{\frac{\gamma_{GG}}{2\beta_0}}-
\eta^{\frac{\gamma_{\gamma\gamma}}{2\beta_0}}
\right) {\widetilde  C}_{8G}(m_W)\non
&&-\frac{\gamma_{G\gamma}}{\gamma_{GG}-\gamma_{\gamma\gamma}}
\left(\eta^{\frac{\gamma_{GG}}{2\beta_0}}
-\eta^{\frac{\gamma_{\gamma\gamma}}{2\beta_0}}\right)
\frac{\beta_{\gamma i}}{2\beta_0}
\hat{V}_{il}^{-1}\frac{1}{a_l-\frac{\gamma_{GG}}{2\beta_0}}
{\hat V}_{lk}{\widetilde C}_k^{(0)}(1)\non
&&+\hat{V}_{il}^{-1}\left[\frac{\beta_{\gamma i}}{2\beta_0}+
\frac{\gamma_{G\gamma}}{2\beta_0}\frac{\beta_{G i}}{2\beta_0}
\frac{1}{a_l-\frac{\gamma_{GG}}{2\beta_0}}\right]
\frac{\eta^{a_l}-\eta^{\frac{\gamma_{\gamma\gamma}}{2\beta_0}}}
{a_l-\frac{\gamma_{\gamma\gamma}}{2\beta_0}}
{\hat V}_{lk}{\widetilde C}_k^{(0)}(1).
\label{c7mu}
\eea

For ${\cal C}_9^{(0)}(\eta)$, the corresponding solution is
\bea
{\cal C}_9^{(0)}(\eta)&=&
\frac{z_{sb}}{K_{tb}K_{ts}^*}\left[
\frac{\pi}{\alpha_{em}}(-1+4\sin^2\theta_W)
+\frac{\pi}{\alpha_s(m_W)}
\sum _{l=1}^{10} \bar{p}_l (\eta^{a_l}-1)\right]
+\frac{\pi}{\alpha_s(m_W)}
\sum_{l=1}^{10}p_l(\eta^{a_l}-1).
\eea
where 
\bea
p_l&=&\sum_{i=1}^{10}\frac{\gamma^{(0)}_{i,11}}{\beta_0}
\hat{V}^{-1}_{il}\hat{V}_{l2},\non
\bar{p}_l&=&-\sum_{i=1}^{10}\frac{\gamma^{(0)}_{i,11}}{\beta_0}
\hat{V}^{-1}_{il}\left[{\hat V}_{l2}
+\frac{1}{6}{\hat V}_{l3}+\frac{2}{3}\sin^2\theta_W{\hat V}_{l7}+
\frac{2}{3}\sin^2\theta_W{\hat V}_{l9}\right].
\eea

To obtain the scheme independent one-loop matrix element of
${\cal O}_9$ in VQM, calculation up to next-to-leading order (NLO) 
is necessary, as the case of SM. 
To obtain ${\cal C}_9(\mu)$ , 
we frist present the NLO calculations  for ${\tilde C}_i(\mu)$  as 
\bea
{\widetilde C}_i^{(1)}(\eta)&=&{\hat V}^{-1}_{il}
\eta^{a_l}{\hat V}_{lk}{\widetilde C}_k^{(1)}(1)
              +{\hat V}^{-1}_{il}
\left[{\hat V}\frac{\gamma^{(1)T}}{2\beta_0}{\hat V}^{-1}\right]_{lj}
\frac{\eta^{a_j-1}-\eta^{a_l}}{a_j-a_l-1}{\hat V}_{jk}
{\widetilde C}_k^{(0)}(1)\non
&&
+\frac{\beta_1}{\beta_0}{\hat V}^{-1}_{il}a_l\left(
\eta^{a_l-1}-\eta^{a_l}\right)
{\hat V}_{lk}{\widetilde C}_k^{(0)}(1).
\label{cb1i}
\eea
The solution of $C_9(\eta)$ in NDR scheme is 
\bea
{\cal C}_9(\eta)&&=\frac{z_{sb}}{K_{tb}K_{ts}^*}\left[
\frac{\pi}{\alpha_{em}}(-1+4\sin^2\theta_W)
+\frac{\pi}{\alpha_s(m_W)}
\sum _{l=1}^{10} \bar{p}_l (\eta^{a_l}-1)\right]
+\frac{\pi}{\alpha_s(m_W)}
\sum_{l=1}^{10}p_l(\eta^{a_l}-1)\non
&&+\left[\frac{C(x_t)-B(x_t)}{\sin^2\theta_W}-4C(x_t)-D(x_t)
+\frac{4}{9}(1+\xi)\right]\non
&&+\sum_l\left[\left(\eta^{a_l}-1\right)r_l+(s_l+q_lE(x_t))
\left(\eta^{a_l+1}-1\right)\right].
\eea
The numerical results for parameters $a_l,p_l,\bar{p}_l,s_l,q_l$ are  
follows: 
\bea
a_l&=      &(-0.8994,-0.5217,-0.4230, 0.1457, 0.2609, 0.4087,0.1304,-1.4034),\non
p_l&=      &(0.1648,  0.2424, 0.1384, -0.0073,-0.3941,0.0433, 0,        0),\non
\bar{p}_l&=&(-0.0248, 1.0004, 0.0853,  0.0207,1.6602,-0.0690,0.5223,-1.6974),\non
s_l&=&(-0.3554, -0.3579, -0.3617,0.0072,-0.2009,0.0490,0,0),\non
q_l&=&(-0.2701,  0,      0.0918,,0.0059,    0, 0.0318,0,0),\non
r_l^{NDR}&=&(-0.0292,-0.1960, 0.1328,-0.1858,0.8997, -0.2011,0,0).
\eea
One can check that in case of $z_{sb}=0$, the results are 
the same as those in SM \cite{Misiak93}. 

Note four-quark operators ${\cal Q}_{1-10}$ can contribute
to one-loop matrix element of ${\cal O}_9$. 
Defining the effective coefficient ${\cal C}_9^{eff}$ as
\bea
{\cal C}_9^{eff}<sl^+l^-|{\cal O}_9|b>\equiv 
{\cal C}_9(\eta)<sl^+l^-|{\cal O}_9|b>+
\sum_{i=1}^{10}C_i^{(0)}(\eta)<sl^+l^-|{\cal Q}_i|b>,
\label{c9eff}
\eea
we obtain $C_9^{eff}(\eta)$ in VQM as 
\bea
C_9^{eff}&=&\frac{z_{sb}}{K_{tb}K_{ts}^*}\left[
\frac{\pi}{\alpha_{em}}(-1+4\sin^2\theta_W)
+\frac{\pi}{\alpha(m_W)}\sum_l\bar{p}_l(\eta^{a_l+1}-1)\right]\non
&&+\frac{\pi}{\alpha_s(m_W)}\sum_l p_l(\eta^{a_l+1}-1)
+\left[\frac{C(x_t)-B(x_t)}{\sin^2\theta_W}-4C(x_t)-D(x_t)
+\frac{4}{9}\right]\non
&&+\sum_l\left[\left(\eta^{a_l}-1\right)r^{NDR}_l+(s_l+q_lE(x_t))
\left(\eta^{a_l+1}-1\right)\right]\non
&&+h(\frac{m_c^2}{m_b^2},s)T_9^{c(0)}+h(1,s)T_9^{b(0)}
-\frac{1}{2}h(0,s)\left[{\tilde C}^{(0)}_3+3{\tilde C}^{(0)}_4\right]
+\frac{9}{2}\left[3{\tilde C}^{(0)}_3+{\tilde C}^{(0)}_4
+3{\tilde C}^{(0)}_5+{\tilde C}^{(0)}_6\right],
\eea
where $s$ is dilepton mass squared normalized by $m_b^2$, and
\bea
T_9^{c(0)}&=&3{\tilde C}_1^{(0)}+{\tilde C}^{(0)}_2
+3{\tilde C}^{(0)}_3+{\tilde C}^{(0)}_4+3{\tilde C}^{(0)}_5
+{\tilde C}^{(0)}_6,\non
T_9^{b(0)}&=&-\frac{1}{2}\left[4{\tilde C}^{(0)}_3+4{\tilde C}^{(0)}_4
+3{\tilde C}^{(0)}_5+{\tilde C}^{(0)}_6
\right].
\eea

The function $h(z,s)$ which includes light quark-antiquark pair contributions
has an expression as 
\bea
h(z,s)&=&-\frac{8}{9}\ln\frac{m_b}{\mu}-\frac{8}{9}\ln z+\frac{8}{27}
      -\frac{4}{9}x-\frac{2}{9}(2+x)|1-x|^{\frac{1}{2}}\non
     &&\times \left\{\begin{array}{ll}
    \ln\left|\frac{\sqrt{1-x}+1}{\sqrt{1-x}-1}\right|-i\pi
     & for\ x\equiv 4\frac{z^2}{s}<1,\\
     2 \arctan\frac{1}{\sqrt{x-1}} 
     & for\ x\equiv 4\frac{z^2}{s}>1.\\
     \end{array}\right.    
\eea
Now we consider the long distance contributions to
${\cal C}_9^{eff}$. Apart from $J/\Psi$ family, 
resonance $\phi$ should be added to long distance
part of ${\cal C}^{eff}_9$ due to tree level $Z\bar{s}b$
coupling.  Similar to previous analysis \cite{Ali97}, 
the non-perturbative contributions can be 
parameterized as 
\bea
Y_{Res}(s)&=&\frac{3\pi}{\alpha_{em}^2}
\left[T^{s(0)}\frac{\Gamma[\phi\to\ell^+\ell^-]m_{\phi}}
{m^2_{\phi}-sm_B^2-m_{\phi}\Gamma_{\phi}}
+T^{c(0)}_9\kappa\sum\limits_{i=1}^6
\frac{\Gamma[\Psi(is)\to\ell^+\ell^-]m_{\Psi(is)}}
{m^2_{\Psi(is)}-sm_B^2-i
m_{\Psi(is)}\Gamma_{\Psi(is)}}
\right],\non
{\cal C}_9^{eff}&&\to {\cal C}_9^{eff}+Y_{Res}(s).
\eea
Here $\kappa=2.3$ is a phenomenological factor,
$|T^{c(0)}_9\kappa|$ can be fixed from the $J/\Psi, 
\Psi^\prime$ data. $T^{s(0)}_9$ and $T^{c(0)}_9$ correspond to
the coefficients of $h(\frac{m_s^2}{m_b^2},s)$ and 
$h(\frac{m_c^2}{m_b^2},s)$ in expression of ${\cal C}_9^{eff}$
respectively, $T^{s(0)}_9=T^{b(0)}$.
Since experimental measurement
for $B\to X_s\phi$  is not available so far, we 
set the phenomenological factor for  $\phi$ to be unity in our numerical 
calculation.

\subsection{Constraints on $z_{sb}$ from  inclusive decays 
$B\to X_s\ell^+\ell^-$}

With all Wilson coefficients at $b$ quark mass scale ready,
the invariant dilepton distribution for 
$B\to X_s\ell^+\ell^-$ can be expressed in terms of the effective    
Wilson coefficients defined above as
\bea
\frac{d\Gamma(B\to X_s l^+l^-)}{ds}&=&
\left(\frac{\alpha_{em}}{4\pi}\right)^2 
\frac{G_F^2m_b^5|K_{ts}^*K_{tb}|^2}{48\pi^3}(1-s)^2R_0
\label{brbsll}
\eea
where 
\bea
R_0&=&4\left(1+\frac{2}{s}\right)
|{\widetilde C}^{eff}_{7\gamma}|^2
+(1+2s)\left(|{\cal C}_9^{eff}|^2+|{\cal C}_{10}|^2\right)
+12Re({\widetilde C}^{eff}_{7\gamma}{\cal C}_9^{eff*}),
\eea
where ${\widetilde C}^{eff}_{7\gamma}$ is determined by 
Eq. (\ref{c7mu}) using the vectors $\beta_\gamma$ and 
$\beta_G$ calculated in HV scheme.

Now we constrain 
the parameter $z_{sb}$ from $B\to X_sl^+l^-$.
The branching ratio of $ B \to X_s l^+ l^-$ depends
on the tree level FCNC coupling $z_{sb}$ and  $|K_{tb}K_{ts}|$.
$|K_{tb}K_{ts}|$ can be written in terms of  $|K_{cb} K_{cs}|$
and $z_{sb}$. In fact, 
 from the quadrangle in $ b\bar{s}$ sector, Fig. \ref{XX},
one can derive the relation,
\bea
|K_{tb}K_{ts}^*|=|K_{cb}K_{cs}^*|
+|K_{ub}K_{us}^*|\cos \theta_3+|z_{sb}|\cos\theta, \non
\theta=arg\left(\frac{z_{sb}}{K_{tb}K_{ts}^*}\right),\ \ \
\theta_3=arg\left(-\frac{K_{ub}K_{us}^*}{K_{tb}K_{ts}^*}\right).
\label{zsbvtbts}
\eea
We use Eq.\ (\ref{zsbvtbts}) to determine $|K_{tb}K_{ts}|$ in Eq.\ (\ref{brbsll}).
Then, from the branching ratios of $B \to X_s l^+ l^-$,
one can derive the constraints on $|z_{sb}|$ and $\theta$.
In the relation Eq.\ (\ref{zsbvtbts}), for  $|K_{cb} K_{cs}|$, we have used the
experimental values shown in Table \ref{Tab1}.  
$|K_{ub} K_{us}|$ in Eq.\ (\ref{zsbvtbts}) was neglected in the numerical calculation.
Integrating Eq.(\ref{brbsll}), we  exclude the resonances $J/\Psi,\Psi'$ 
contributions by using the same cuts as experiments\cite{Belle03} 
$
0.20<|m_{l^+l^-}-m_{J/\Psi}|<0.35,\quad 
0.15<|m_{l^+l^-}-m_{\Psi^\prime}|<0.30.
$
The  corresponding $2\sigma$ experimental bounds on 
the size of $z_{sb}$ and phase $\theta$  
from experimental measurement for 
$B\to X_sl^+l^-$ are displayed in Fig.{\ref{Fig7}.
From this figure, we obtain
\bea	
|z_{sb}|\le 1.40\times 10^{-3} \ (95\%\ \  C.L.).     
\label{boundzsb}
\eea 

One can see clearly that 
the absolute value of the allowed phase $\theta$
becomes smaller as the size of $z_{sb}$ increases.

\begin{table}[hbt]
\caption{Input parameters are used in calculation. All masses in 
unit of GeV.}
\begin{tabular}{c  c | c c}
$\alpha_s(m_Z)$  & 0.119 & $\alpha_{em}$      &   1/133\\
$m_Z$ &  91.19        &$\lambda_2$  & $0.12\ GeV^2$ \\
$m_W$ &  80.41        &$Br^{ex}(B\to X_ce\nu)$ & 0.104\\
$m_b^{pole}$ & 4.8    &$|K_{us}|$ & $0.2196\pm 0.0026$\\
$m_t^{pole}$ & $173.8\pm 5$ & $|K_{cs}|$&$0.97 \pm 0.11$\\ 
$m_c/m_b$    & $0.29\pm 0.2$& $|K_{ub}|$& $(3.6\pm 0.7)\times 10^{-3}$\\
$m_s/m_b$    & 0.02 & $|K_{cb}|$ & $(4.12\pm 0.2)\times 10^{-2}$\\	
$\sin^2\theta_W$ &   0.231&$G_F$ & $1.166\times 10^{-5}\ GeV^{-2}$ 
\end{tabular}
\label{Tab1}
\end{table}

\begin{figure}[htb]
\centerline{\epsfig{file=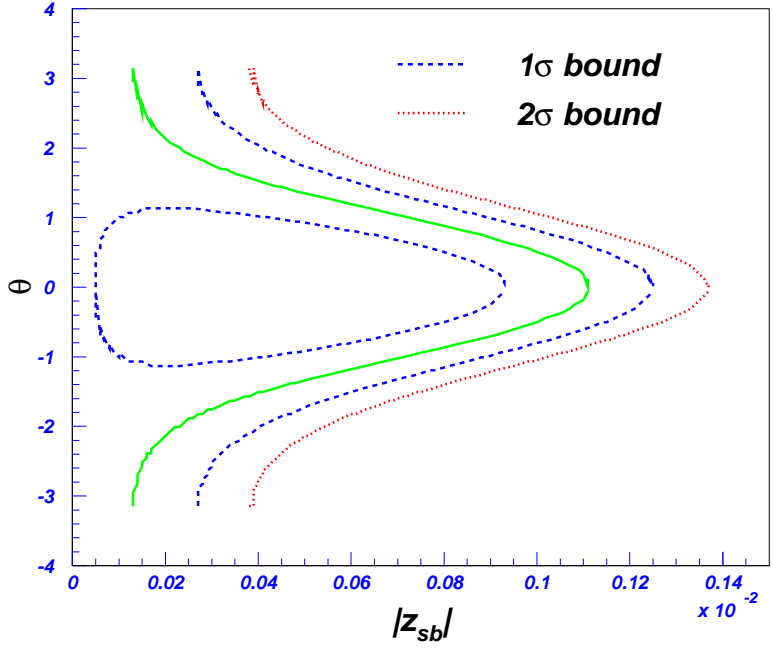,width=12cm}}
\caption{ The ($|z_{sb}|,\theta$) contour in $ Scenario\ A$
 constrained by $Br^{ex}(B\to X_sl^+l^-)$.
The dashed, dotted lines correspond to 
$1\sigma$ and $2\sigma$ experimental bounds of $B\to X_sl^+l^-$
in  (\ref{exbsll}), respectively. The solid line denotes  
the experimental central value $6.2\times 10^{-6}$. 
The region between dot lines is allowed at $1\sigma$ level.}
\label{Fig7}
\end{figure}

\section{Some predictions for  $B\to X_s\ell^+\ell^-$ and 
$B\to X_s\gamma$} 
  
In this section, subject to the constraints on $z_{sb}$ from  
$B\to X_sl^+l^-$, we will present predictions for 
the invariant dilepton mass distribution,
FB asymmetry, the zero point of FB asymmetry
of $B\to X_sl^+l^-$, and the branching ratio of $B\to X_s\gamma$.

\begin{figure}[htb]
\centerline{\epsfig{file=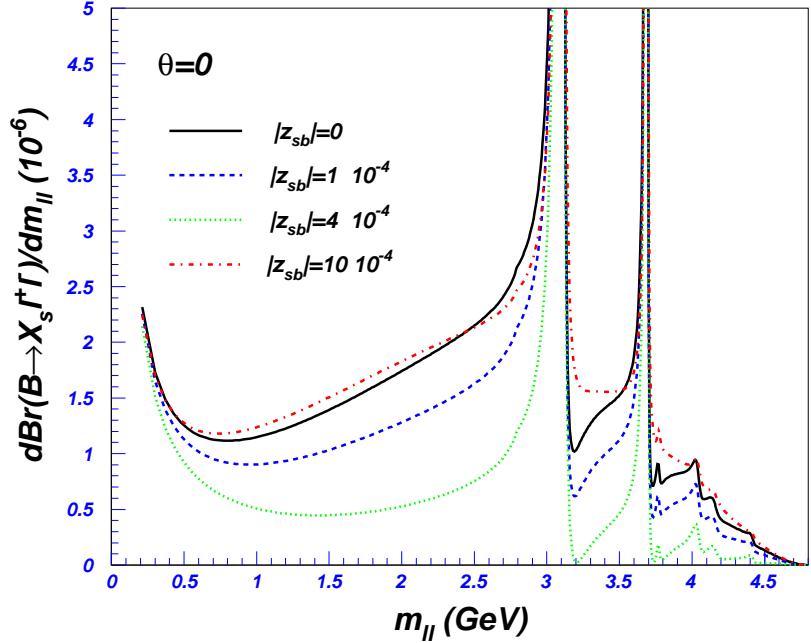,width=12cm}}
\caption{ The differential rate of 
$B\to X_sl^+l^-$ normalized by $Br(B\to X_c l\nu)$ as a function of 
dilepton invariant mass.}
\label{dbdsll}
\end{figure} 

Firstly, we consider the invariant dilepton 
mass distribution of $B\to X_sl^+l^-$.
According to the study in 
subsection \ref{sec:bsll}, it is expected to be sensitive to 
the parameter $z_{sb}$. This feature is  shown  in Fig.\ref{dbdsll}.

In addition to the differential and total branching ratio,
the forward-backward (FB) asymmetry can provide crucial information
on new physics. The normalized FB asymmetry distribution is defined
as  
\bea
\bar{A}_{FB}&=&\frac{\int d{\hat \theta}
\frac{d^2\Gamma( b\to s l^+l^-)}
{dsd\cos{\hat \theta}}sign(\cos{\hat \theta})}
{\int d{\hat \theta}
\frac{d^2\Gamma(b\to s l^+l^-)}{dsd\cos{\hat \theta}}}
=-\frac{3}{R_0}Re[(s{\cal C}^{eff}_9+2{\tilde C}_{7\gamma})
{\cal C}_{10}^*].
\eea
Thus, the  zero of the FB asymmetry $s_0$ in VQM is determined by
equation
\bea
Re\left[\left(s_0{\cal  C}^{eff}_9+
2{\widetilde C}^{eff}_{7\gamma}\right)
{\cal C}_{10}^*\right]=0.
\eea
\begin{figure}[htb]
\centerline{\epsfig{file=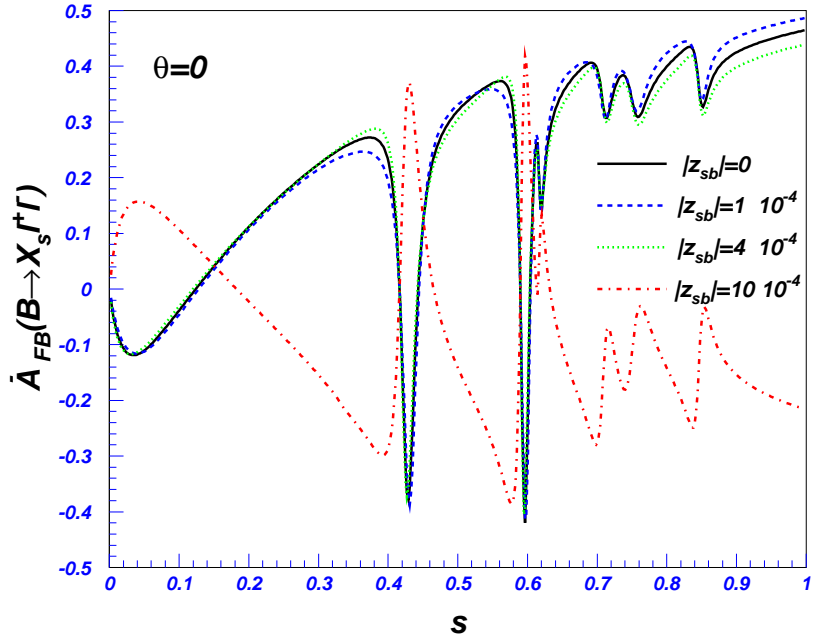,width=12cm}}
\caption{The FB asymmetry of $B\to X_sl^+l^-$ distribution.}
\label{Fig8}
\end{figure}

It is very interesting to analyze how 
the zero of the FB asymmetry ($s_0$) is modified in VQM. 
Unlike to the case of SM where ${\cal C}_{10}$ is real,
the coefficient ${\cal C}_{10}$ is complex generally in VQM. 
Furthermore, the contributions to ${\cal C}_{9,10}$
from  tree level FCNC diagram
\bea
|{\cal C}_{10}^{Tree}(m_W)|
=\frac{\pi}{\alpha_{em}}
\left|\frac{z_{sb}}{K_{tb}K_{ts}^*}\right|
\gg |{\cal C}_{9}^{Tree}(m_W)|
=\frac{\pi}{\alpha_{em}}\left|\frac{z_{sb}}{K_{tb}K_{ts}^*}\right|
(1-4\sin^2\theta_W),
\eea
indicate that ${\cal C}_{10}$ can have 
large  imaginary part. Therefore, $s_0$  in VQM 
will have large deviation from  that in SM.

As an illustration, we plot the FB asymmetry as a function of $s$
in Fig.\ \ref{Fig8} corresponding to different size of $z_{sb}$. 
To show how the zero point $s_0$  sensitive to both the size of $z_{sb}$ 
and phase $\theta$, we display the dependence of zero point of FB 
asymmetry on $z_{sb}$ in Fig. \ref{Fig9}. 
Figs.\ref{Fig8} and \ref{Fig9} indicate that, 
subject to the experimental 
measurement for branching ratio of $B\to X_sl^+l^-$, 
FB asymmetry distribution and the zero point of 
FB asymmetry are  very sensitive to 
the parameters $z_{sb}$ and phase $\theta$, especially 
in the region $0.6\times 10^{-3}<|z_{sb}|<1.2\times 10^{-3}.$    
\begin{figure}[htb]
\centerline{\epsfig{file=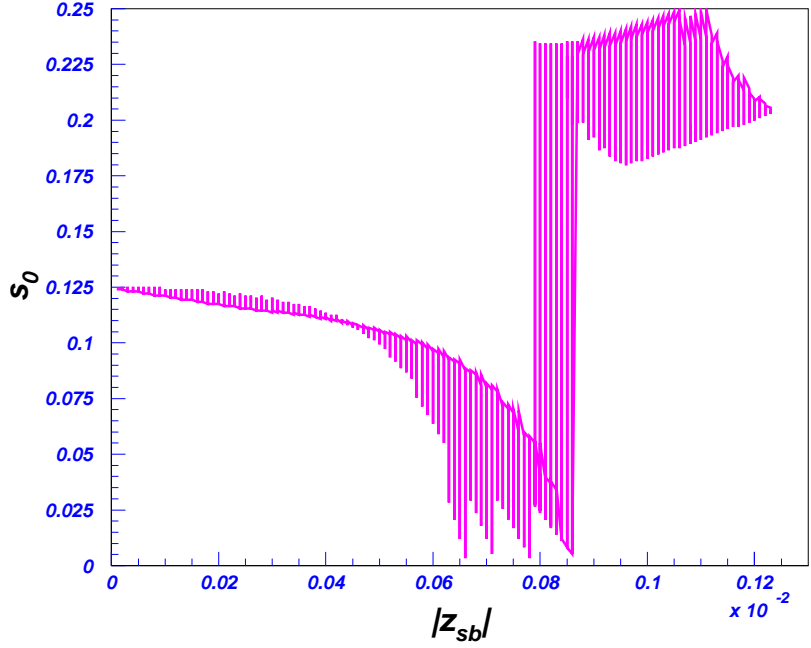,width=12cm}}
\caption{The $(|z_{sb}|, s_0)$ contour in VQM 
subject to the $1\sigma$ bounds of $Br^{ex}(B\to X_sl^+l^-)$.
For specified $|z_{sb}|$, the phase $\theta$ effect on 
$s_0$ is also shown.} 
\label{Fig9}
\end{figure}

Now we study the radiative decay $B\to X_s\gamma$.
The Branching ratio of  $B\to X_s\gamma$  can be expressed as 
\bea
{\cal B}r(B\to X_s\gamma)&=&
{\cal B}r^{ex}(B\to X_ce\bar{\nu}_e)
R_{quark}({\delta}^{max})
\left[1-6\left(1-\frac{(1-z)^4}{g(z)}\right)
\lambda_2\right],
\eea
where 
$z=m_c^2/m_b^2,\ g(z)=1-8z+8z^3-z^4-12z^2\ln z$ is the phase space factor.
${\delta}^{max}=0.99$ \cite{Chetyrkin97}.
The  $\lambda_2$-dependent term comes from the nonperturbative corrections
to the semileptonic and radiative B meson decay rates, and 
$\lambda_2=(m^2_{B^*}-m_B^2)/4\simeq 0.12\ GeV^2$.
The ratio $R_{quark}$ is defined as 
\bea
R_{quark}({\delta})&=&
\frac{\Gamma[b\to X_s\gamma]^{E_\gamma>(1-\delta)m_b/2}}
{\Gamma[b\to X_ce\bar{\nu}_e]}
=\frac{6\alpha_{em}}{\pi g(z)}
\left|\frac{K_{tb}K^*_{ts}}{K_{cb}}\right|^2F(|D|^2+A),
\label{rquark}
\eea 
where  function $F$ is related to the next-to-leading QCD corrections 
to the semileptonic decay \cite{Cabibbo78},
\bea
F&=&
\left(1-\frac{8}{3}\frac{\alpha_s(m_b)}{\pi}\right)
\left[1-\frac{2}{3}\frac{\alpha_s(m_b)}{\pi}\frac{h(z)}{g(z)}\right]^{-1}.
\eea
The $|D|^2$ term in (\ref{rquark}) 
stands for the contribution of $b\to s\gamma$ 
while $A$ term which is cutoff $\delta$ dependent, includes 
the virtual and Bremsstrahlung correction to $b\to X_s\gamma$
\cite{Chetyrkin97}. They can be written in terms of 
Wilson coefficients as 
\bea
D&=&{\widetilde C}_{7\gamma}(\mu_b)
+\frac{\alpha_s(\mu_b)}{4\pi}\sum\limits_i^8{\widetilde C}_i^{(0)}(\mu_b)
\left[\hat{r}_i+\hat{\gamma}_{i7}^{(0)}
\ln\frac{m_b}{\mu_b}\right],\non
A&=&\left(e^{-\alpha_s(\mu_b)/3\pi(7+2\ln\delta)\ln\delta}-1\right)
|{\widetilde C}_{7\gamma}^{(0)}(\mu_b)|^2
+\frac{\alpha_s(\mu_b)}{\pi}
\sum\limits_{i\leq j}{\widetilde C}_i^{(0)}(\mu_b) 
{\widetilde C}_j^{(0)}(\mu_b)f_{ij}(\delta),
\eea
where $\hat{\gamma}_{i7}^{(0)}=\beta_{\gamma i}$ for $i=1-6$,
$\hat{\gamma}_{77}^{(0)}=\gamma_{\gamma\gamma}$ and
$\hat{\gamma}_{87}^{(0)}=\gamma_{G\gamma}$. 
The explicit expressions of $\hat{r}_i$,and $f_{ij}(\delta)$ can be found 
in Ref. \cite{Chetyrkin97}. 

\begin{figure}[htb]
\centerline{\epsfig{file=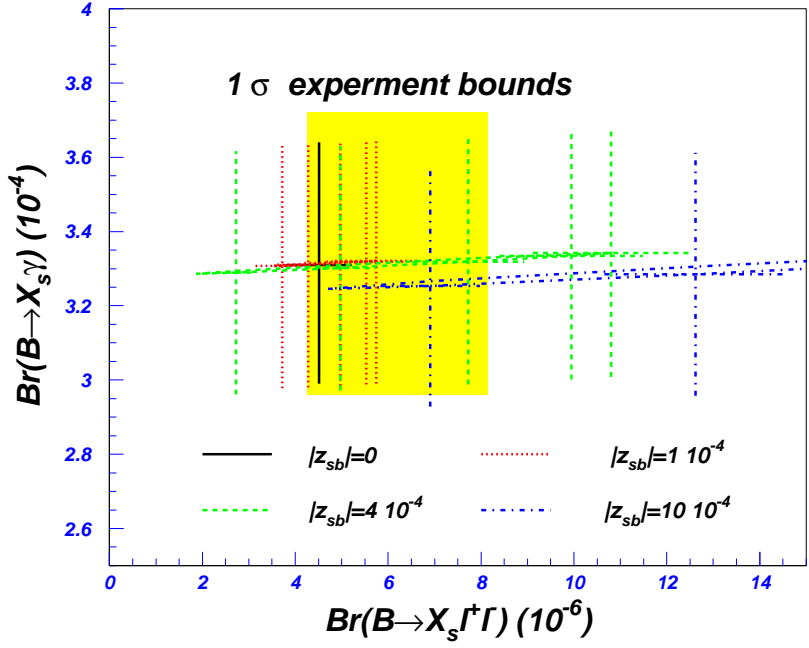,width=12cm}}
\caption{
The correlation of the branching ratios of  
$B\to X_s\gamma$ and $B\to X_sl^+l^-$ predicted in VQM.
The solid, dashed, dot and  dot-dashed lines 
in the figure stand for different $|z_{sb}|$.
Each $|z_{sb}|$ corresponds to 
one of the five specified 
$\theta=(-\pi,-\frac{3}{5}\pi, -\frac{1}{5}\pi, 
\frac{1}{5}\pi,\frac{3}{5}\pi)$.
The experimental $1\sigma$ bounds denoted as rectangular box 
are taken from (\ref{exbsll}) and (\ref{bsrex}).} 
\label{Fig6}  
\end{figure} 

Fig. \ref{Fig6} shows correlation of the branching ratio of 
$B\to X_s\gamma$ with that of $B\to X_sl^+l^-$ in VQM.
It is obvious from the figure that 
$Br(B\to X_s\gamma)$ is not so sensitive to the phase, which
is not the case for  $Br(B\to X_sl^+l^-)$. Within the experimental bounds of
$B\to X_sl^+l^-$ , corresponding  branching ratio of 
$B\to X_s\gamma$ predicted in VQM  is consistent with
the experiment measurement \cite{Jssop02} 
\bea
{\cal B}r^{ex}(B\to X_s\gamma)=(3.34\pm 0.38)\times 10^{-4}.
\label{bsrex}
\eea

\section{ $B\to K\pi$ decays }

Having established constraints on $b\to s$ FCNC from $b\to sll$ and 
$b\to s\gamma$, we evaluate $B\to K\pi$ decay rates. The large 
electroweak penguin contribution to $b\to s$ transition has been
suggested in the present data of B factories. 
Because in VQM, the tree level FCNC may generate the electroweak penguin 
contribution, the purpose of this section is that  with FCNC constrained
by rare decays $b\to sl^+l^-$, we study whether the large electroweak 
penguin contribution in $B\to K\pi$ can be explained or not. 
The  correlation between di-leptonic decay and non-leptonic decays is characteristic feature of the present model. 

We start with the following effective Hamiltonian:
\bea
H_{eff}=\frac{4G_F}{\sqrt{2}}\left\{(\xi_c (C^*_1 \bar{\cal Q}_1
+C_2^* \bar{\cal Q}_2)
+ \xi_u (C^*_1 \bar{\cal Q}^u_1
+C^*_2 \bar{\cal Q}^u_2)
-\xi_t \sum_{i=1}^{10} (C^*_i \bar{\cal Q}_i) \right\}.
\eea
The operators $\bar{\cal Q}_i$ is the charge conjugation of ${\cal Q}_i$ 
defined in (\ref{operators}), $\xi_i=K_{is}K_{ib}^*$,  and 
\bea
\bar{\cal Q}^u_1=(\bar{b}_i\gamma^\mu L u_j)(\bar{u}_j\gamma_\mu L{s}_i),
\ \ 
\bar{\cal Q}^u_2=(\bar{b}_i\gamma^\mu L u_i)(\bar{u}_j\gamma_\mu L{s}_j).
\eea
Here we must keep the $\xi_u$ term because there are tree level contribution
to $B\to K\pi$ decays.
As seen from the effective Hamiltonian, tree-level Z FCNC generates the 
electroweak penguin contribution which is not suppressed by $\alpha_{em}$.
In terms of isospin amplitudes, the electroweak penguin operators 
contribute to both $\Delta I=0$ and $\Delta I=1$ components. 
Because for  $\Delta I=0$  amplitude, the large strong penguin contribution
is expected, new physics may dominantly contribute to 
$\Delta I=1$  amplitudes. 

Below, we  estimate the $B \to K \pi$ amplitudes within the factorization
approximation. The four  quark operator with the flavor structure
$b\to s\bar{s}{s}$ and  $b\to s\bar{c}{c}$  do not contribute to the processes.
We obtain 
\bea
A(B^+\to K^0\pi^+)&=&iG_F\left\{
\xi_t\left(a_4-\frac{a_{10}}{2}+(2a_6-a_8)R^K\right)
M^{B\pi,K}\right.\non
&&\left.+\left[\xi_u a_2-\xi_t\left(a_4+a_{10}+2(a_6+a_8)R^B\right)\right]
  M^{K\pi,B}\right\},\\
A(B^+\to K^+\pi^0)&=&i\frac{G_F}{\sqrt{2}}\left\{
\left[\xi_u \frac{a_1}{2}-\xi_t\frac{3}{4}(-a_7+a_9)\right]
\sqrt{2}M^{BK,\pi}
\right.\non
&&\left.+\left[a_2\xi_u-\xi_t\left(a_4+a_{10}+2(a_6+a_8)R^K\right)
\right]
M^{B\pi,K}\right.\non
&&\left.-\left[\xi_u a_2-\xi_t\left(a_4+a_{10}+2(a_6+a_8)R^B\right)
\right]M^{K\pi,B}\right\},\\
A(B^0\to K^+\pi^-)&=&iG_F\left\{\left[\xi_u a_2-\xi_t\left(a_4+a_{10}+2(a_6+a_8)R^K\right)\right]
M^{B\pi,K}\right.\non
&&\left.+\xi_t\left(a_4-\frac{a_{10}}{2}+(2a_6-a_8)R^B\right)
M^{K\pi,B}\right\},\\
A(B^0\to K^0\pi^0)&=&i\frac{G_F}{\sqrt{2}}\left\{\left[
\xi_u \frac{a_1}{2}-\xi_t\frac{3}{4}(-a_7+a_9)\right]
\sqrt{2}M^{BK,\pi}\right.\non
&&\left.+\xi_t\left(a_4-\frac{a_{10}}{2}+(2a_6-a_8)R^K\right)
M^{B\pi,K}\right.\non
&&\left.
-\xi_t\left(a_4-\frac{a_{10}}{2}+(2a_6-a_8)R^B\right)
M^{K\pi,B}\right\}
\eea
where $a_{2i-1}=C^*_{2i-1}+ C^*_{2i}/N$, 
$a_{2i}=C^*_{2i}+ C^*_{2i-1}/N$. And 
\bea
R^K&=&\frac{m_K^2}{(m_b-m_u)(m_s+m_u)},\ \ \
R^B=\frac{m_B^2}{(m_b+m_u)(m_s-m_u)}, \non
M^{ab,c}&=&f_c(M_a^2-m_b^2)F_0^{a\to b}(m_c^2).
\eea
In derving the amplitudes, we have used 
\bea
<\pi^0(p^\prime))|(\bar{b}u)_V|B^+(p)>
&=&\left[(p+p^\prime)_\mu-\frac{m_B^2-m_\pi^2}{q^2}q_\mu\right]
F_1^{B^+\to\pi^0}(q^2)
+\frac{m_B^2-m_\pi^2}{q^2}q_\mu F_0^{B^+\to \pi^0}(q^2),\non
<K^+(p^\prime))|(\bar{b}s)_V|B^+(p)>
&=&\left[(p+p^\prime)_\mu-\frac{m_B^2-m_K^2}{q^2}q_\mu\right]F_1^{B\to K}(q^2)
+\frac{m_B^2-m_K^2}{q^2}q_\mu F_0^{B\to K}(q^2),\non
<K^+(-p)\pi^0(p^\prime))|(\bar{u}s)_V|0>
&=&\left[(p+p^\prime)_\mu 
-\frac{m_K^2-m_\pi^2}{q^2}q_\mu\right]F_1^{K\to\pi}(q^2)
+\frac{m_K^2-m_\pi^2}{q^2}q_\mu F_0^{K\to \pi}(q^2),
\eea
where $q=p-p^\prime$, and 
\bea
&& <0|(\bar{s}u)_A|K^+(p_k)>=if_K p_{k\mu},\ \ \ \
 <0|(\bar{b}u)_A|B(p_B)>=if_B p_{B\mu},\non
&& <K^+(p_k)|(\bar{s}u)_A|0>=-if_K p_{k\mu},\ \ 
 <\pi^0(p_\pi)|(\bar{u}u-\bar{d}d)_A|0>=-i \sqrt{2} f_{\pi} p_{\pi\mu}.
\eea
It is easy to see that the isopsin relation
\bea
\sqrt{2}A(B^+\to K^+\pi^0)+A(B^+\to K^0\pi^+)
=\sqrt{2}A(B^0\to K^0\pi^0)+A(B^0\to K^+\pi^-).
\eea
Since there are many uncertainties in calculating $B\to K\pi$ such as 
strong phases and formfactors, instead of the branching ratios of 
$B\to K\pi$,  it is more reasonable to use their ratios to 
constrain the new physics effect. In this work, we use $R_c,\ R_n$ 
and their difference to constrain Z FCNC. $R_c$ and $R_n$ are 
defined as 
\bea
R_c&\equiv& \frac{2Br(B^+\to K^+\pi^0)+2Br(B^-\to K^-\pi^0)}
            {Br(B^+\to K^0\pi^+)+Br(B^-\to \bar{K}^0\pi^-)},\non
R_n&\equiv&\frac{Br(B^0\to K^+\pi^-)+Br(\bar{B}^0\to K^-\pi^+)}
            {2Br(B^0\to K^0\pi^0)+2Br(\bar{B}^0\to \bar{K}^0\pi^0)}.
\eea
\begin{figure}[htb]
\begin{center}
\epsfig{file=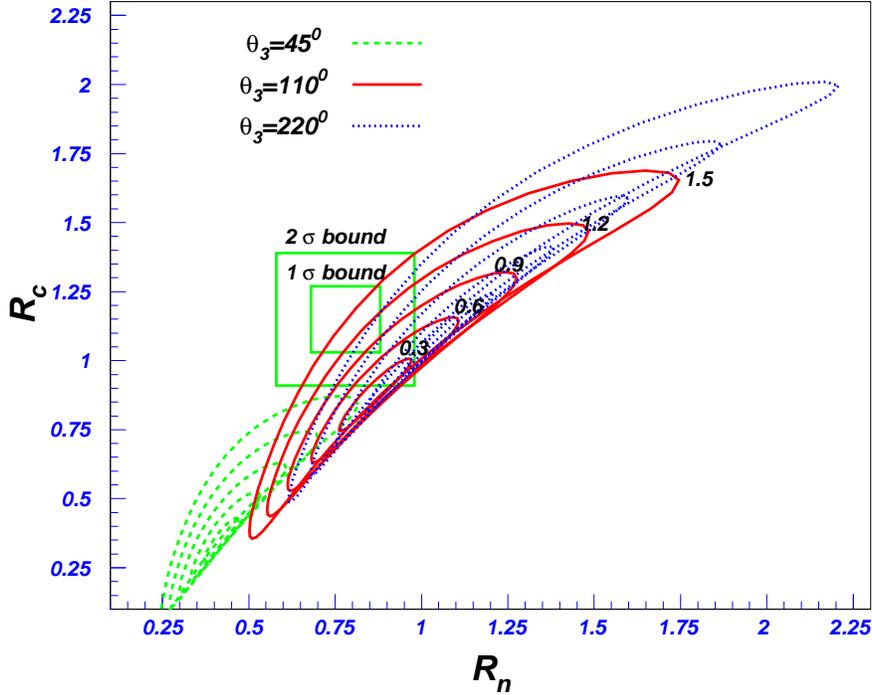,width=12cm}
\end{center}
\caption{Correlation between $R_n$ and $R_c$. 
The dashed, solid and dotted lines denote for the phase
$\theta_3=45^0, 110^0, 220^0$, respectively. 
For each specified phase $\theta_3$, the sizes of  $z_{sb}$
corresponding to contours (inner to outer) are 
$(0.3,0.6,0.9,1.2,1.5)\times 10^{-3}$.}
\label{figA}
\end{figure}
In our numerical calculations, we take large $N$ limit and 
use $f_\pi=132\ GeV$.
Since $q^2=m_K^2, m_\pi^2$ is rather close to the point
$q^2=0$, we neglect the $q^2$ dependence in the formfactors, while 
$F_0^{B\to K}=0.33, \  F_0^{B\to \pi^0}=0.379/\sqrt{2}$.
The SU(3)-breaking effects in the formfactors are also neglected.
The contribution from the terms proportional
to $f_B$ is  also omitted because it works as a suppression 
factor $f_B/M_B$\cite{Yoshikawa03}.

Running quark masses appear in the matrix elements of $(S-P)(S+P)$
penguin operators through the use of the motion equation. 
The running quark masses at $\mu\sim m_b=4.8\ GeV$
scale are given by 
\bea
m_b(m_b) =4.34 GeV,\ \  m_s(m_b)=0.09 GeV, \ \ m_u(m_b)=m_d(m_b)=0.03 GeV.
\eea
The NLO Wilson coefficients in NDR scheme at $m_b$ scale can 
be obtained from (\ref{cb0i}) and (\ref{cb1i}). 

As discussed in Section II, the phases $\theta_3$ and $\theta$ 
are defined in (\ref{zsbvtbts}) and are shown in
 Fig.\ \ref{XX}. We can write $\theta_3$ as,
\bea
\theta_3\simeq  -atan\left(\frac{\eta}{\rho}\right),
\eea
where $ \eta$ and $ \rho$ can be detremined from the measuerments
of CP violation and mixings of $K$  and $B_d$ systems.
By assuming that the effect of tree level FCNC on $K$ and
$B_d$ is small, we can use the standard model values
for  $\eta$ and $\rho$. The  allowed region
of $\theta_3$ is:
\bea
100^0 \le \theta_3 \le 140^0.
\label{theta3}
\eea
If the tree level FCNC contribute to  $K$ and $B_d$ system, 
the allowed region of $\theta_3$ can be significantly changed
from Eq.(\ref{theta3}).
Therefore, in  our numerical calculation, we relax this condition and
scan the 
region for $\theta_3$ from $0^0$ to $360^0$.
Now we can see the physics quantities are deterimined by three inputs:
the size of Z FCNC coupling $|z_{sb}|$, the  phases of $\theta_3$ 
and $\theta$. 
We plot  the $R_n$ and $R_c$ correlation in Fig. \ref{figA}
with three specified values of $\theta_3$.
For given $z_{sb}$, the allowed region changes with $\theta_3$.
For $|z_{sb}|\leq 1.5\times 10^{-3}$ and 
$\theta_3\leq 45^0$, no region is allowed 
by $R_c$ and $R_n$ experimental measurements at $2 \sigma$ level. 
When  $\theta_3=100^0\sim 120^0$,
as shown in the figure,  there are larger allowed regions.

To see this more clearly, in Fig.\ref{figB} we 
show the dependence of $R_c-R_n$ on the  phase 
$\theta_3$ with typical values
 $|z_{sb}|=(1.0, 1.5)\times 10^{-3}$ and $\theta=(-\pi,-\frac{\pi}{2},
0,\frac{\pi}{2},\pi)$.  
The experimental measurements \cite{HFAG}
\bea
R_c=1.15\pm 0.12,\ \ R_n=0.78\pm 0.10, \ \ 
R_c-R_n=0.37\pm 0.16
\eea
at $1\sigma $ and $2 \sigma$ level are also displayed.
\begin{figure}[htb]
\begin{center}
\epsfig{file=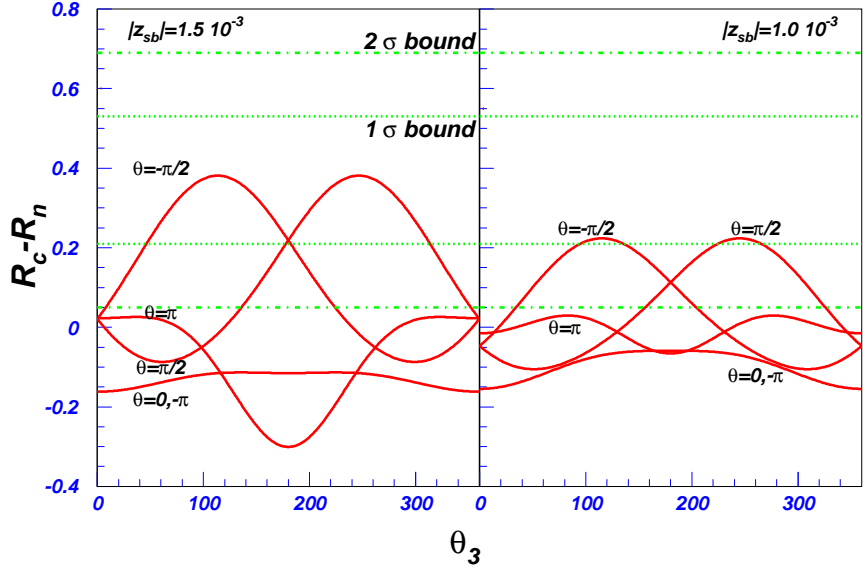,width=12cm}
\end{center}
\caption{$R_c-R_n$ as a funition of the phase $\theta_3$.}
\label{figB}
\end{figure}
From this figure, we note  that $R_c-R_n$ is sensitive to 
the phase $\theta_3$ and $\theta$. 

The constraints on the contour $(|z_{sb}|, \theta)$ from  $R_c-R_n$
are shown in Fig. \ref{figC}.
One can see that, for $\theta_3=110^0$, 
$1\sigma $ experimental bounds restrict $|z_{sb}|$ and $\theta$ as 
$|z_{sb}|>0.9\times 10^{-3}$, $|\theta|\ge 1$.
Corresponding to  $2\sigma $ experimental bounds,  
smaller $|z_{sb}|$ is allowed as $|z_{sb}|>0.2\times 10^{-3}$.

 \begin{figure}[htb]
\begin{center}
\epsfig{file=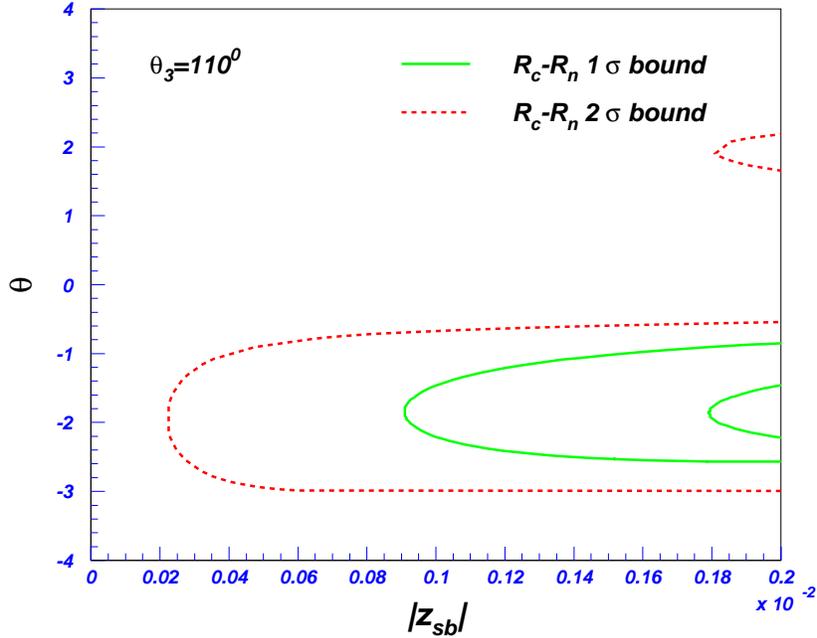,width=11cm}
\end{center}
\caption{The $(|z_{sb}|,\theta)$ contour constrained by $R_c-R_n$ 
at $1\sigma$ and $2\sigma$ level.}
\label{figC}
\end{figure}

Finally, we obtain the allowed regions 
from $R_c-R_n$ and $B\to X_sl^+l^-$ by varying $\theta_3$ as 
a free parameter. The bounds are shown in Fig. \ref{figE}.
At $1\sigma$ level, there are no overlapping region allowd from 
$B\to X_sl^+l^-$ and $B\to K\pi$.
From Fig. \ref{figE}, we obtain
\bea
0.2\times 10^{-3}<|z_{sb}|<1.2 \times 10^{-3}, \ (95\%\ C.\ L.)
\eea
where  large phase $|\theta|$ is favored. 
Now we would like to point out that if 
$\theta_3$ is in the region of 
$[100^0,140^0]$, only the region with negative phase $\theta$ 
in Fig. \ref{figE} is allowed.

\begin{figure}[bth]
\begin{center}
\epsfig{file=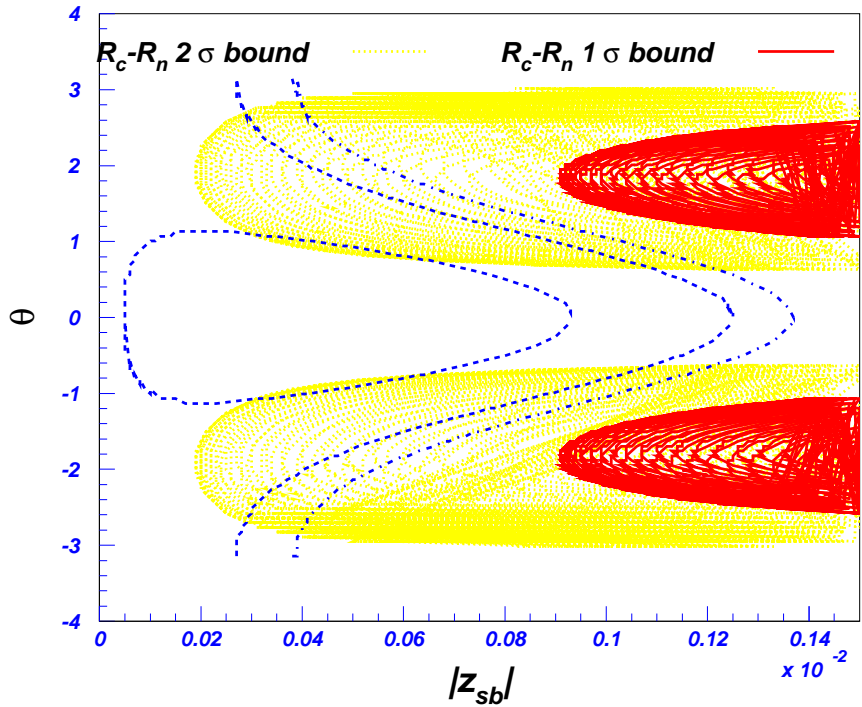,width=12cm}
\end{center}
\caption{The $(|z_{sb}|,\theta)$ contour constrained by combination 
of $B\to X_sl^+l^-$ and $B\to K\pi$  at $1\sigma$ and  $2\sigma$ level.
The solid and dotted lines correspond to 
$1\sigma$ and $2\sigma$ bounds from $R_c-R_n$, respectively. 
The dash-dotted line and dashed lines correspond 
to $2\sigma$ and $1\sigma$ bounds from $B\to X_sl^+l^-$.
The allowed region with $\theta>0$ corresponds to $180^0<\theta_3<360^0$
whereas the allowed region with $\theta<0$ 
corresponds to $0^0<\theta_3<180^0$.}
\label{figE}
\end{figure}

\section{Conclusions}

In this work, we have studied  
the new physics effects on the B meson
decays $B\to X_s\gamma, B\to X_s\ell^+\ell^-$ and $B\to K\pi$ 
in the Vectorlike  Quark Model. The QCD running effect from the mass of 
the down-type vector quark $D$ to weak scale  has been taken
into account.
We extracted rather stringent constraints on the size of 
the tree level Z FCNC coupling and its CP violating phase
from $B\to X_s\ell^+\ell^-$.
Within the bounds,  we investigated  various observables
such as FB asymmetry of $ b \to s l^+ l^-$, the decay rates of
$B \to X_s \gamma$ and $B\to K\pi$. 
We found that the zero crossing 
point of FB  asymmetry $s_0$ of $B\to X_sl^+l^-$ is very sensitive to 
both the size and phase of $z_{sb}$ and can be useful in probing 
the new physics, and both experimental measurements 
for $B\to X_sl^+l^-$ and $B\to K\pi$ decays can be explained
within the framework of vector-like quark model.
The upper bound on the size of $Z$ FCNC comes from 
$B\to X_sl^+l^-$ , while the lower bound, from $B\to K\pi$ decays.    
Considering that the B factories such as BaBar and Belle are running,  
measurements  for inclusive and exclusive B decays 
with high precision are expected. Therefore, 
the VQM will be tested in near future.  

\section*{Acknowledgement} 
We would like to thank Profs. Y.Y. Keum and C.D.\ L\"{u} 
for useful discussions.
The work of T. Morozumi and Z. Xiong is supported by 
the Grant-in-Aid for JSPS Fellows (No. 1400230400).
 The work of T.Y. is 
supported by 21st Century COE Program of Nagoya University provided 
by JSPS (15COEG01).
\appendix
\section{}
In this appendix, we present all anomalous dimension matrices
needed in solving the  Wilson coefficients. 

(1) For the mixings of operators $Q^H_{LR},\ R^{1,H}_L,\ R^{2,H}_L$, 
we obtain
\bea 
\gamma=\left(\begin{array}{ccc}
            7& 0 & 0\\
            0&7& 0\\
            0& 0&7\\
           \end{array}\right).
\label{DM1}
\eea
We also obtain the same matrix for 
$Q^{\chi}_{LR},\ R^{1,\chi}_L,\ R^{2,\chi}_L$ as Eq.(\ref{DM1}).
For the mixings  among the operators 
$O_{LR}^i \ (i=1,2,3)$ and $P_L^{1,A},\ P_L^2$, 
we obtain the following result:
\bea
\gamma=\frac{g_s^2}{8\pi^2}\left(\begin{array}{cccccccc} 
       \frac{20}{3} & 1 & -2 & 0 &0 & 0& 0 &0 \\
           -8 & \frac{2}{3}& \frac{4}{3}& 0 &0 & 0& 0&0 \\
         0&0 &\frac{16}{3}& 0 &0 & 0& 0 &0\\
         6&2&-1&\frac{2}{3}&2&-2&-2&0\\
         4&\frac{3}{2}&0&-\frac{113}{36}&\frac{137}{18}
          &-\frac{113}{36}&-\frac{4}{3}&\frac{9}{4}\\
         2&1&1&-2&2&\frac{2}{3}&-2&0\\
         0&\frac{1}{2}&2&-\frac{113}{36}&\frac{89}{18}&-\frac{113}{36}&
          \frac{4}{3}&\frac{9}{4}\\
        \end{array}
    \right),
\label{DM2}
\eea
which is in agreement with Ref.\cite{Gao95}. 

(2) The $10\times 10$ one-loop anomalous dimension 
matrix  among ${\cal Q}_{1-10}$ is given by
\cite{Buras94}:
\bea
\gamma^{(0)}=\left(\begin{array}{cccccccccc} 
       -2 & 6 & 0 & 0 &0 & 0& 0 & 0&0&0\\
        6 & -2& \frac{-2}{9}&\frac{2}{3}&\frac{-2}{9}&\frac{2}{3}&0&0&&0\\
         0&0 &\frac{-22}{9}& \frac{22}{3}&\frac{-4}{9}&\frac{4}{3}&0&0&0&0\\
         0&0 &6-\frac{2f}{9}& -2+\frac{2f}{3}&\frac{-2f}{9}&\frac{2f}{3} &0&0&&0\\
         0&0 &0 &0 &2& -6&0&0&0&0\\
         0&0 &-\frac{2f}{9}&\frac{2f}{3}&
\frac{-2f}{9}&-16+\frac{2f}{3} &0&0&0&0\\
         0&0 &0 &0 &0&0&2 &-6&0&0\\
0&0&\frac{-2(u-d/2)}{9}&\frac{2(u-d/2)}{3}
&\frac{-2(u-d/2)}{9}&\frac{2(u-d/2)}{3}&0&-16&0&0\\
0&0&\frac{2}{9}&\frac{-2}{3}&\frac{2}{9}&\frac{-2}{3}&0&0&-2&6\\
0&0&\frac{-2(u-d/2)}{9}&\frac{2(u-d/2)}{3}
&\frac{-2(u-d/2)}{9}&\frac{2(u-d/2)}{3}&0&0&6&-2\\
        \end{array}
    \right)
\eea
where the color number $N=3$ is used.
For $b\to sl^+l^-$, $f=u+d,\  u=2, d=3$. $\gamma^{(0)}$ is scheme 
independent.

In NDR scheme the $10\times 10$ two-loop anomalous dimension 
matrix among the ${\cal Q}_{1-10}$  is given by 
\bea
\gamma^{NDR,(1)}=\begin{array}{c}
\left(\begin{array}{ccccc} 
\frac{-21}{2}-\frac{2f}{9}& \frac{7}{2}+\frac{2f}{3}&\frac{79}{9} 
&\frac{-7}{3}  & \frac{-65}{9}\\
 \frac{7}{2}+\frac{2f}{3}& \frac{-21}{2}-\frac{2f}{9}&
\frac{-202}{243} &\frac{1354}{81}  & \frac{-1192}{243}\\
0&0&\frac{-5911}{486}+\frac{71f}{9} &\frac{5983}{162}+\frac{f}{3} & 
\frac{-2384}{243}-\frac{71f}{9}\\
0&0&\frac{379}{18}+\frac{56f}{243} &\frac{-91}{6}+\frac{808f}{81} & 
\frac{-130}{9}-\frac{502f}{243}\\
0&0&\frac{-61f}{9} &\frac{-11f}{3}& 
\frac{71}{3}+\frac{61f}{9}\\
0&0&\frac{-682f}{243} &\frac{106f}{81}& 
\frac{-225}{2}+\frac{1676f}{243}\\
0&0&\frac{-61(u-d/2)}{9}&\frac{-11(u-d/2)}{3}&
\frac{83(u-d/2)}{9}\\
0&0&\frac{-682(u-d/2)}{243}&\frac{106(u-d/2)}{81}&
\frac{704(u-d/2)}{243}\\
0&0&\frac{202}{243}+\frac{73(u-d/2)}{9}&\frac{-1354}{81}-\frac{(u-d/2)}{3}&
\frac{1192}{243}-\frac{71(u-d/2)}{9}\\
0&0&\frac{-79}{9}-\frac{106(u-d/2)}{243}&\frac{7}{3}+\frac{826(u-d/2)}{81}&
\frac{65}{9}-\frac{502(u-d/2)}{243}\\
        \end{array}\right.\\
\left.\begin{array}{ccccc}
-\frac{7}{3}&0&0&0&0\\
\frac{904}{81}&0&0&0&0\\
\frac{1808}{81}-\frac{f}{3}&0&0&0&0\\
\frac{-14}{3}+\frac{646f}{81}&0&0&0&0\\
-99+\frac{11f}{3}&0&0&0&0\\
\frac{-1343}{6}+\frac{1348f}{81}&0&0&0&0\\
\frac{-11(u-d/2)}{3}&\frac{71}{3}-\frac{22f}{9}&-99+\frac{22f}{3}&0&0\\
\frac{736(u-d/2)}{81}
&\frac{-225}{3}+4f&-\frac{1343}{6}+\frac{68f}{9}&0&0\\
-\frac{904}{81}-\frac{(u-d/2)}{3}
&0&0&\frac{-21}{2}-\frac{2f}{9}&\frac{7}{2}+\frac{2f}{3}\\
\frac{7}{3}+\frac{646(u-d/2)}{81}
&0&0&\frac{7}{2}+\frac{2f}{3}&\frac{-21}{2}-\frac{2f}{9}\\
  \end{array}\right).
\end{array}
\eea

(3) The mixing entries among ${\cal Q}_{7\gamma,8G}$ and  
${\cal Q}_{1-10}$  are follows:
\bea
\gamma_{\gamma \gamma}=\frac{32}{3},\ \  \gamma_{G\gamma}=
-\frac{32}{9}, \ \  \gamma_{GG}=\frac{28}{3}.
\eea
In HV scheme, the entries of  $\beta_{\gamma,G}^{7-10}$ can be extracted 
from $\beta_{\gamma,G}^{3-6}$ \cite{Ciuchini94} by substituting 
\bea
1\to \frac{3}{2}e_d, \ \ u\to \frac{3}{2}e_uu, \ \ d\to \frac{3}{2}e_dd,
\label{c16710}
\eea
and thus, they are given by 
\bea 
\beta_\gamma&=&\left(0,\frac{416}{81},-\frac{464}{81},
(\frac{416}{81}u-\frac{232}{81}d),\frac{32}{9},
-(\frac{448}{81}u-\frac{200}{81}d),
-\frac{16}{9}, -(\frac{448}{81}u^2+\frac{100}{81}d^2),
\frac{232}{81},-(\frac{416}{81}u^2-\frac{116}{81}d^2)\right),\non
\beta_G&=&\left(3,\frac{70}{27},\frac{140}{27}+3f,
6+\frac{70}{27}f,-\frac{14}{3}-3f,-4-\frac{119}{27}f, 
\frac{7}{3}-3(u-\frac{d}{2}), 2-\frac{119(u-d/2)}{27},\right.\non
&&\left. 
-\frac{70}{27}+3(u-\frac{d}{2}),
-3+\frac{70(u-d/2)}{27}\right),
\eea

(4) The mixing entries among ${\cal C}_{9}$ and  
${\cal Q}_{1-10}$  are follows \cite{Buras94} with same 
substitution in (\ref{c16710}):
\bea
\gamma^{(0)}_{i,11}&=&\left(-\frac{16}{3},\frac{-16}{9},
\frac{-16}{3}(u-\frac{d}{2}-\frac{1}{3}),\frac{-16}{9}(u-\frac{d}{2}-3),
\frac{-16}{3}(u-\frac{d}{2}),\right.\non
&&\left.-\frac{16}{9}(u-\frac{d}{2}),-(\frac{16}{3}u^2+\frac{4}{3}d^2),
-(\frac{16}{9}u^2+\frac{4}{9}d^2),
-(\frac{16}{3}u^2+\frac{4}{3}d^2+\frac{8}{9}),
-(\frac{16}{9}u^2+\frac{4}{9}d^2+\frac{8}{3})
\right),\\
\gamma^{NDR,(1)}_{i,11}&=&\left(-\frac{64}{3}, 
  \frac{1600}{243}, -\frac{3712}{243} - \frac{64(u-d/2)}{3}, 
 \frac{ 64}{3} -\left(\frac{2240}{243}u+\frac{512}{243} d\right), -\frac{64(u-d/2)}{3},
 \frac{3520}{243}u-\frac{3392}{243}d,\right.\non
&&\left.-(\frac{64}{3}u^2+\frac{16}{3}d^2),
 (\frac{3520}{243}u^2+\frac{1696}{243} d^2),
 \frac{1856}{243}-(\frac{64}{3}u^2+\frac{16}{3}d^2), 
  -\frac{32}{3} -(\frac{2240}{81}u^2-\frac{256}{243} d^2)\right),\\
\label{gammai11} 
\gamma^{(0)}_{11,11}&=&-2\beta_0, \ \ 
\gamma^{NDR,(1)}_{11,11}=-2\beta_1,
\eea
where $\beta_0=11-2f/3,\beta_1=102-38f/3$. 

\section{}
Some one-loop functions needed in our calculations are follows.

(1) For the calculation of the initial values of 
${\widetilde C}_{1-8}$ and ${\cal C}_{9,10}$ at $m_W$ scale: 
\bea
A(x)&=&-\frac{8x^3+5x^2-7x}{12(1-x_t)^3}
+\frac{x^2(2-3x)}{2(1-x)^4}\ln x,\non
B(x)&=&-\frac{x}{4(x-1)}+\frac{x}{4(x-1)^2}\ln x,\non
C(x)&=&\frac{x^2-6x}{8(x-1)}+\frac{3x^2+2x}{8(x-1)^2}\ln x,\non
D(x)&=&-\frac{19x^3-25x^2}{36(x-1)^3}
-\frac{3x^4-30x^3+54x^2-32x+8}{18(x-1)^4}\ln x,\non
E(x)&=&-\frac{2}{3}\ln x +\frac{x(18-11x-x^2)}{12(1-x^3)},\non
F(x)&=&-\frac{x^3-5x^2+2x}{4(1-x)^3}
+\frac{3x^2}{2(1-x)^4}\ln x.
\eea

(2) For the calculation of one-loop diagrams with down-type
vector quark $D$:  
\bea
f_D^Z(x)&=&-\frac{5x^2+5x-4}{72(x-1)^3}
+\frac{x(2x-1)}{12(x-1)^4}\ln x,\non
f_D^H(x)&=&-e_dx
\left[\frac{7x^2-29x+16}{48(x-1)^3}+\frac{3x-2}{8(x-1)^4}
\ln x\right],\non
f_D^\chi(x)&=&e_dx\left[\frac{5x^2-19x+20}{48(x-1)^3}
+\frac{x-2}{8(x-1)^4}\ln x\right],
\label{funcD}
\eea

\end{document}